\documentclass{article}
 \usepackage{natbib}
 \usepackage{bm}
\usepackage{tikz}
\usepackage{booktabs}
\usepackage{threeparttable}
\usepackage{adjustbox}
\usepackage{colortbl}
 \usepackage{amsmath}
\usepackage{hyperref}
\usepackage{longtable}
\usepackage{algorithm}
\usepackage{algorithmic}
\usepackage{multirow}
\usepackage{threeparttablex}
\usepackage{makecell}
\usepackage{dsfont}
\usepackage{tabularx}
\usepackage{subcaption} 
\usepackage{enumitem} 
\usepackage{lmodern} 
\usepackage{authblk}
\usepackage{amsfonts}
\usepackage{natbib} 
\usepackage[a4paper,left=2cm, right=2cm, top=1.50cm, bottom=1.50cm]{geometry}
\newcommand{\ba}{{\textbf{a}}}
\newcommand{\bb}{{\textbf{b}}}

\newcommand{\mat}[1]{\bm{#1}} 
\newcommand{\obs}{{\text{obs}}} 
\newcommand{\mis}{{\text{mis}}} 
\newcommand{\mar}{{\text{MAR}}} 
\newcommand{\mnar}{{\text{MNAR}}}

\hypersetup{
    colorlinks=true,
    linkcolor=blue!50!black,
    citecolor=blue!50!black,
    urlcolor=blue!50!black,
    pdfborder={0 0 0} 
}

\newtheorem{diagnostic}{Diagnostic}

\title{Sensitivity analysis method in the presence of a  missing not at random ordinal independent variable}

\tiny{\author[1,2]{\small Abdoulaye Dioni}
\author[1,3]{Alexandre Bureau}
\author[1,4]{Lynne Moore}
\author[1,2]{Aida Eslami}
\affil[1]{ Département de Médecine Sociale et Préventive
  Université Laval, Québec, Canada}
\affil[2]{ Institut Universitaire de Cardiologie
 et de Pneumologie de Québec, Québec, Canada}
 \affil[3]{Centre de Recherche CERVO, Québec, Canada}
 \affil[4]{Centre de recherche du CHU de Québec,  Québec, Canada}}
\date{}

\begin{document}
\maketitle
\vspace{2cm}

\section*{Abstract}

Data analysis often encounters missing data, which can result in inaccurate conclusions, especially when it comes to ordinal variables. In trauma data, the Glasgow Coma Scale is useful for assessing the level of consciousness. This score is  often missing in patients who are intubated or under sedation upon arrival at the hospital, and those with normal reactivity without head injury, suggesting a Missing Not At Random (MNAR) mechanism. 
The problem with MNAR is the absence of a definitive analysis. While sensitivity analysis is often recommended, practical limitations sometimes restrict the analysis to a basic comparison between results under Missing Completely At Random (MCAR) and Missing At Random (MAR) assumptions, disregarding MNAR plausibility. Our objective is to propose a flexible and accessible sensitivity analysis method in the presence of a MNAR ordinal independent variable. The method is inspired by the sensitivity analysis approach proposed by Leurent {\textit{et al}}. (2018)   for a continuous response variable. We propose an extension  for an independent ordinal variable. The method is evaluated on simulated data before being applied to Pan-Canadian trauma data from April 2013 to March 2018.
The simulation shows that MNAR estimates are less biased than MAR estimates and more precise than complete case analysis (CC) estimates. The  confidence intervals coverage rates are relatively better for MNAR estimates than CC and MAR estimates. In the application, it is observed that the Glasgow Coma Scale is significant under MNAR, unlike MCAR and MAR assumptions.\\

\textbf{Keywords:} Glasgow Coma Scale, Missing Not At Random, ordinal variable, sensitivity analysis.

\newpage

\section{Introduction}\label{Introduction:chp3}

Data analysis in social sciences and public health faces challenges due to the categorical variables with possible predefined orders, correlated observations and missing values. 
Considering the structure of correlated observations, such as hierarchical data organized at multiple levels (e.g., patients within hospitals, hospitals within provinces), it is crucial for obtaining precise estimates and accurate variance components \citep{Goldstein2011,Jiang2021,mcculloch2008}.
By definition, missing values are unobserved values that would be meaningful for the analysis if observed \citep{Little2020,NationalResearchCouncil2010}. In other words, ignoring missing values during analysis is likely to introduce potential bias, loss of information, and reduced precision \citep{buuren2018,fitzmaurice2008,Gomer2021}.

The taxonomy of missing data distinguishes three classes of mechanisms: (i) Missing Completely At Random (MCAR), if the probability of data being missing is independent of observed and unobserved data, (ii) Missing At Random (MAR), if the probability of data being missing depends only on the observed data  and (iii) Missing Not At Random (MNAR), if the probability of data being missing depends on unobserved data and possibly observed data \citep{Gomer2021,Little2020, molenberghs2005}. In particular, the MNAR mechanism is the most problematic as it requires additional untestable assumptions about the missing data process \citep{molenberghs2014,carpenter2023, leurent2018}. 

Under MCAR, several ad hoc solutions exist, such as complete-case analysis (CC), available-case analysis, mean or mode imputation  \citep{Yan2009,Seaman2013, Little2020}.  Some authors consider the MCAR assumption strong for real data, recommending the MAR assumption as the starting point for handling missing data \citep{molenberghs2005,hox2011}. 

Under  MAR, there are also methods such as Maximum Likelihood, Bayesian methods, Inverse Probability Weighting (IPW), and Multiple Imputation (MI) \citep{Kang2015,enders2023,Moore2009, Tsiatis2006,Seaman2013}. In particular, MI is more attractive due to the availability and accessibility of analysis packages  \citep{buuren2011, Quartagno2023,gelman2011,kalpourtzi2024,molenberghs2005,enders2022}. 
In practice, it is challenging to dismiss the  MNAR possibility \citep{NationalResearchCouncil2010,Moreno-Betancur2016,oberman2024}. 

Under MNAR and in the non-hierarchical context, there are two main models namely  Pattern-Mixture Models (PMM) and Selection Models (SM) \citep{Little2020,molenberghs2014}. PMM stratifies the data according to the configurations of missing data \citep{Little2020,Yuan2009}. Furthermore, PMM is more suitable in situations where there is no need to fully model the mechanism of missing data \citep{molenberghs2014, Little1993}. In contrast, SM factorizes   the joint distribution of the partially observed variable and the missing data process into a marginal distribution of the partially observed variable and the missing data process conditionally on other variables \citep{thijs2002,molenberghs2005,Yuan2009}. In the hierarchical context, SM and  PMM correspond to Mixed-Effect Selection Models (MESM) and Mixed-Effect Pattern-Mixture Models (MEPMM) respectively \citep{molenberghs2014, enders2022,Yuan2009}. However,  a definitive analysis under MNAR does not exist, some authors suggest conducting sensitivity analysis to study the deviation from MAR \citep{Little1993, hammon2023, thijs2002,NationalResearchCouncil2010}. 

Sensitivity analysis can be defined as analyses in which several statistical models are considered simultaneously and/or where a statistical model is further investigated using specialized tools, such as diagnostic measures \citep{molenberghs2014,mallinckrodt2020}. 
A flexible approach to conducting sensitivity analysis is the PMM model using MI method \citep{ Moreno-Betancur2016, buuren2018, leurent2018}.   
Some authors propose a  \texttt{SensMice} package in the \texttt{R} software as an extension of the \texttt{mice} package for conducting sensitivity analysis \citep{Resseguier2011}.  
However, \texttt{SensMice} does not handle the hierarchical context and  the partially observed  ordinal variables.  
Research in sensitivity analysis methods is active and relatively recent \citep{molenberghs2014, carpenter2021,carpenter:sen:2007}.

To our knowledge, no method for sensitivity analysis  of an independent ordinal variable with missing only two specific categories has been explored.
This paper aims to develop a  sensitivity analysis method under MNAR,  focusing on ordinal independent variable based on MI.   
The method, is evaluated  on simulated data and  Pan-Canadian trauma data from April 2013 to March 2018 to determine variation in mortality across provinces. The article is structured as follows. 
Section \ref{nota:chap3} introduces the method. Section \ref{method:chap3} outlines our sensitivity analysis method. Section \ref{simulation:study:chap3} covers the simulation study.  Section \ref{simulation:result:chp3} presents the simulation results. Section \ref{Application:chap3} describes the Pan-Canadian trauma data and the main findings.  Section \ref{Discussion:chap3} discusses the main findings.

\noindent In this work, we focus on estimating the regression coefficients linking the outcome \(Y\) to the covariates, including a partially observed ordinal covariate \(X_1\). These coefficients represent conditional effects. We do not target the residual variance parameter, which plays no role in the interpretation of the estimand under consideration. While alternative estimands—such as marginal effects obtained by standardizing over the covariate distribution—may be relevant in other contexts, they are not the focus of the present analysis.

\section{Method}\label{nota:delta:chap3}

\subsection{Notation and assumptions} \label{nota:chap3}
Let $(Y, \mat{X})$ denote a dataset consisting of $n$ independent observations. The outcome variable is denoted \( Y = (y_1, \dots, y_n)^\top \), and \( \mat{X} = (X_1, X_2, \dots, X_J) \) is a matrix of covariates. We define \( \mat{X}_{\{-1\}} = (X_2, \dots, X_J) \) as the submatrix excluding the first covariate \( X_1 \), and \( \mat{x}_{\{i,-1\}} \) denotes the $i$th row of this submatrix. We assume that \( X_1 \) is an ordinal covariate with \( K > 2 \) ordered categories, and that its values are partially observed, subject to a Missing Not at Random (MNAR) mechanism. Specifically, the probability of missingness may depend on the unobserved value of \( X_1 \), making standard missing data techniques inappropriate without additional assumptions. We let \( X_1 = \{ X_1^{\obs}, X_1^{\mis} \} \) denote the partition of observed and missing values. The missingness indicator is defined as \( R_i = 1 \) if \( X_{1i} \) is observed, and \( R_i = 0 \) otherwise. All other variables, including \( Y \) and \( \mat{X}_{\{-1\}} \), are assumed to be fully observed.
In clustered settings, a group identifier variable \( clus \) is used to indicate the cluster membership for each unit. Throughout this manuscript, we are primarily interested in estimating the regression coefficients linking \( Y \) to the covariates, and we examine how inference may be biased under an MNAR mechanism affecting \( X_1 \). We now introduce the ordinal regression model that serves as the theoretical foundation for our sensitivity analysis procedure.

\subsection{Ordinal regression model based on a latent variable}\label{sec:ordinal_latent_model}

To model the distribution of the ordinal covariate \( X_1 \), we adopt a latent variable formulation frequently used in the literature \citep{venables2002, christensen2022}. In this framework, we assume the existence of a continuous latent variable \( \theta_i \) such that:
\[
\theta_i = \beta^y Y_i + \mat{x}_{\{i,-1\}}^\top \beta^x + \varepsilon_i,
\]
where \( \beta^y \) and \( \beta^x \) are regression parameters, and \( \varepsilon_i \sim \mathcal{N}(0,1) \) is a standard normal error term, corresponding to a probit link function. The observed ordinal value \( X_{1i} \) is then derived from \( \theta_i \) using a set of increasing threshold parameters \( \zeta = (\zeta_0, \zeta_1, \dots, \zeta_K) \), with \( \zeta_0 = -\infty \) and \( \zeta_K = +\infty \). Specifically,
\[
X_{1i} = k \quad \text{if } \theta_i \in (\zeta_{k-1}, \zeta_k], \quad \text{for } k = 1, \dots, K.
\]
Under this cumulative probit model, the conditional probability that \( X_{1i} \) falls in category \( k \) is given by:
\[
P(X_{1i} = k \mid Y_i, \mat{x}_{\{i,-1\}}) = \Phi(\zeta_k - \eta_i) - \Phi(\zeta_{k-1} - \eta_i),
\]
where \( \eta_i = \beta^y Y_i + \mat{x}_{\{i,-1\}}^\top \beta^x \), and \( \Phi(\cdot) \) denotes the cumulative distribution function  of the standard normal distribution.

This ordinal regression model serves as the conceptual foundation for imputing missing values of \( X_1 \) under a MAR assumption, using standard implementations such as proportional odds models (e.g., \texttt{mice}) in non-hierarchical settings \citep{buuren2011} or latent normal regression models (e.g., \texttt{jomo}) in hierarchical or multilevel contexts \citep{Quartagno2023}. In practice, \texttt{mice} preserves the ordinal structure of the imputed variable, whereas \texttt{jomo} treats it as categorical by imputing the underlying latent variable. To maintain coherence with our framework, we systematically reconvert the imputed values from \texttt{jomo} into ordinal categories based on the estimated thresholds.
In our sensitivity analysis, we extend this ordinal model to explore plausible MNAR mechanisms by adjusting the threshold parameters. These deviations and their implications are formalized and motivated in the following section using Directed Acyclic Graphs (DAGs), with respect to their impact on the regression estimand of interest.

\subsection{Missing data assumptions using DAGs}

Directed Acyclic Graphs (DAGs) are valuable tools for visualizing, encoding, and communicating assumptions about the missing data mechanism \citep{thoemmes2015, enders2023}. Building on the latent-variable-based ordinal regression model presented previously, this section illustrates how different assumptions about the missingness of $X_1$---namely MCAR, MAR, and MNAR---impact the identifiability of the estimand, particularly the regression coefficients linking $Y$ to $\mat{X}$.

We adopt a graphical notation adapted from \citet{thoemmes2015}, introducing the proxy variable $X^{\star}_1$, defined as $X^{\star}_1 = X_1^{\obs}$ if $R = 1$ and $X^{\star}_1 = X_1^{\mis}$ if $R = 0$. The variables $\epsilon_{R}$, $\epsilon_{X_1}$, $\epsilon_{Y}$, and $\epsilon_{\mat{X}_{\{-1\}}}$ represent unspecified sources of variation. Observed variables ($Y$, $R$, $\mat{X}_{\{-1\}}$) are represented by solid rectangles, while partially observed variables like $X_1$ are shown with dashed borders. Dashed arrows indicate unmeasured or untestable pathways.

\textbf{i. MCAR (Figure \ref{dag-MCAR}).} Here, $X^{\star}_1$ is a collider between $R$ and $X_1$ along the path $R \rightarrow X^{\star}_1 \leftarrow X_1$, which is blocked. $R$ is independent of all other variables, both observed and unobserved. As such, missingness is completely at random, and standard complete-case or imputation methods yield unbiased estimates.

\textbf{ii. MAR (Figure \ref{dag-MAR}).} In this case, $R$ is d-connected to $X_1$ and $Y$ through $\mat{X}_{\{-1\}}$, via the paths $R \rightarrow \mat{X}_{\{-1\}} \rightarrow X_1$ and $R \rightarrow \mat{X}_{\{-1\}} \rightarrow Y$. Conditioning on $\mat{X}_{\{-1\}}$ blocks these paths, satisfying the MAR assumption and justifying imputation models that condition only on fully observed covariates.

\textbf{iii. MNAR (Figure \ref{dag-MNAR}).} The presence of the path $R \leftarrow X_1$ indicates a direct dependence between missingness and unobserved values. This path cannot be blocked by conditioning on observed variables, thus violating MAR. In such cases, unbiased estimation requires either auxiliary data or sensitivity analyses that encode explicit assumptions about the missing data process \citep{Little1993, molenberghs2014}. Given the difficulty of obtaining external data---e.g., due to dropout or mortality \citep{Little2020}---a pragmatic solution is to apply a delta-adjustment approach to quantify the potential bias due to MNAR. Our method is based on this principle and is detailed in the next section.

\vspace{0.5cm}

\begin{figure}[H]
\centering
\begin{minipage}[b]{0.23\textwidth}
\centering
\resizebox{\textwidth}{!}{
\begin{tikzpicture}[>=stealth, inner sep = 0.5mm,  text width = 1.1cm, text centered]
\node at ( 0,0) [rectangle,draw] (b1) {$Y$ };
\node at ( 0,-1.5) [rectangle,draw] (b2) {$\mat{X}_{\{-1\}}$ };
\node at ( 2,0) [rectangle,draw, dashed] (c1) {$X_1$ };
\node at ( 2,-1.5) [rectangle,draw] (c2) {$X^{\star}_1$ };
\node at ( 2,-3) [rectangle,draw] (c3) {$R$ };
\node [above of=b1]  (E1) {$\epsilon_{Y}$};
\node [below of=b2]  (E2) {$\epsilon_{\mat{X}_{\{-1\}}}$};
\node [above of=c1]  (Ec1) {$\epsilon_{X_1}$};
\node [below of=c3]  (Ec2) {$\epsilon_{R}$};
\draw[->] (b2) -- (b1);
\draw[->] (c1) -- (b1);
\draw[->] (b2) -- (c1);
\draw[->] (c1) -- (c2);
\draw[->] (c3) -- (c2);
\draw[->] (E1) -- (b1);
\draw[->] (E2) -- (b2);
\draw[->] (Ec1) -- (c1);
\draw[->] (Ec2) -- (c3);
\end{tikzpicture}
}
\caption{DAG when $X_1$ is MCAR}
\label{dag-MCAR}
\end{minipage}
\hfill
\begin{minipage}[b]{0.23\textwidth}
\centering
\resizebox{\textwidth}{!}{
\begin{tikzpicture}[>=stealth, inner sep = 0.5mm,  text width = 1.1cm, text centered]
\node at ( 0,0) [rectangle,draw] (b1) {$Y$ };
\node at ( 0,-1.5) [rectangle,draw] (b2) {$\mat{X}_{\{-1\}}$ };
\node at ( 2,0) [rectangle,draw, dashed] (c1) {$X_1$ };
\node at ( 2,-1.5) [rectangle,draw] (c2) {$X^{\star}_1$ };
\node at ( 2,-3) [rectangle,draw] (c3) {$R$ };
\node [above of=b1]  (E1) {$\epsilon_{Y}$};
\node [below of=b2]  (E2) {$\epsilon_{\mat{X}_{\{-1\}}}$};
\node [above of=c1]  (Ec1) {$\epsilon_{X_1}$};
\node [below of=c3]  (Ec2) {$\epsilon_{R}$};
\draw[->] (b2) -- (b1);
\draw[->] (c1) -- (b1);
\draw[->] (b2) -- (c1);
\draw[->] (c1) -- (c2);
\draw[->] (c3) -- (c2);
\draw[->] (c3) -- (b2);
\draw[->] (E1) -- (b1);
\draw[->] (E2) -- (b2);
\draw[->] (Ec1) -- (c1);
\draw[->] (Ec2) -- (c3);
\end{tikzpicture}
}
\caption{DAG when $X_1$ is MAR}
\label{dag-MAR}
\end{minipage}
\hfill
\begin{minipage}[b]{0.30\textwidth}
\centering
\resizebox{\textwidth}{!}{
\begin{tikzpicture}[>=stealth, inner sep = 0.5mm,  text width = 1.1cm, text centered]
\node at ( 0,0) [rectangle,draw] (b1) {$Y$ };
\node at ( 0,-1.5) [rectangle,draw] (b2) {$\mat{X}_{\{-1\}}$ };
\node at ( 2,0) [rectangle,draw, dashed] (c1) {$X_1$ };
\node at ( 2,-1.5) [rectangle,draw] (c2) {$X^{\star}_1$ };
\node at ( 2,-3) [rectangle,draw] (c3) {$R$ };
\node [above of=b1]  (E1) {$\epsilon_{Y}$};
\node [below of=b2]  (E2) {$\epsilon_{\mat{X}_{\{-1\}}}$};
\node [above of=c1]  (Ec1) {$\epsilon_{X_1}$};
\node [below of=c3]  (Ec2) {$\epsilon_{R}$};
\draw[->] (b2) -- (b1);
\draw[->] (c1) -- (b1);
\draw[->] (b2) -- (c1);
\draw[->] (c1) -- (c2);
\draw[->] (c3) -- (c2);
\draw[->] (c3) -- (b2);
\draw[<-] (c3) to[in = 360, out=360] (c1);
\draw[->] (E1) -- (b1);
\draw[->] (E2) -- (b2);
\draw[->] (Ec1) -- (c1);
\draw[->] (Ec2) -- (c3);
\end{tikzpicture}
}
\caption{DAG when \\ $X_1$ is MNAR}
\label{dag-MNAR}
\end{minipage}
\end{figure}

\subsection{The delta adjustment method}\label{delta:method:chap3}

The method known as delta adjustment is grounded in the pattern-mixture model (PMM) framework, where the missing data mechanism does not need to be fully modeled \citep{molenberghs2014, Little1993}. As discussed in the previous section, the MNAR mechanism implies that the missingness depends on unobserved data. In such cases, sensitivity analysis can be conducted by specifying plausible assumptions external to the observed data. This approach enables an evaluation of the robustness of the estimated quantities of interest---such as regression coefficients linking \(Y\) to covariates---under various assumptions regarding the missing data mechanism. Specifically, it does not aim to recover the true data-generating process, but rather to explore how inference might vary across plausible MNAR scenarios informed by expert judgment. Under the delta adjustment approach, multiple imputation is performed in three steps: (i) imputation under the MAR assumption, (ii) modification of the imputed values to reflect plausible MNAR scenarios, and (iii) analysis and combination of results using Rubin's rules \citep{Little2020}.

To illustrate this approach, \citet{leurent2018} propose a method for partially observed continuous \emph{response} variables \(Y\) in a non-hierarchical setting. Specifically, they modify the MAR-based imputation via $Y^{\mnar} = Y^{\mar} + c$ where \(Y^{\mar}\) represents the imputed value under the MAR assumption, \(Y^{\mnar}\) its adjusted version under MNAR, and \(c\) a scalar sensitivity parameter reflecting the assumed degree of departure from MAR. This parameter is not estimated from the data but specified externally based on expert knowledge or scenario exploration. For example, values such as \(c = -1\), \(0\), or \(+1\) may represent conservative, neutral, or anti-conservative deviations from MAR, respectively. A multiplicative version, \(Y^{\mnar} = Y^{\mar} \times c\), is also sometimes used, though it can be counterintuitive when \(Y\) is on a log scale. Several other authors have described similar strategies for continuous outcomes, including \citet{buuren2018}, \citet{Little2020}, and \citet{Rezvan2018}. However, these approaches are limited to continuous response variables. In contrast, our work focuses on a setting where the variable with missing values is a covariate \(X_1\), which is ordinal, and the outcome \(Y\) may be continuous, binary, or categorical. In this case, the sensitivity adjustment does not operate directly on \(Y\), but rather on the imputed values of \(X_1\), by perturbing the latent intercepts of the ordinal model described earlier. 

Importantly, our sensitivity parameter \(\delta = (\delta_1, \dots, \delta_{K-1})\) is a vector that induces category-specific shifts in the threshold parameters. This generalization allows for a flexible and interpretable class of MNAR deviations tailored to ordinal explanatory variables. In the next section, we formally introduce this extension and demonstrate how it can be implemented to perform sensitivity analysis in the presence of partially observed ordinal covariates.

\section{Sensitivity analysis method}\label{method:chap3}

Subsection \ref{method:step} presents our extension, and Subsection \ref{method:justify:chp3} provides its justification framework.

\subsection{Method Steps}\label{method:step}

\textbf{1. Impute missing data under MAR.} \\
Multiple imputation is performed under the MAR assumption. We use the \texttt{R} package \texttt{mice} (version 3.16.0) in non-hierarchical contexts, and \texttt{jomo} (version 2.7.6) in hierarchical settings \citep{buuren2011, Quartagno2023}. The \texttt{mice} package implements a fully conditional specification (FCS), where each incomplete variable is imputed using its own model. Ordinal variables are handled via proportional odds logistic regression \citep{buuren2018}. In contrast, \texttt{jomo} relies on a joint modeling approach under multivariate normality, where ordinal variables are treated as latent continuous variables and imputed accordingly \citep{carpenter2023}.
This step yields an imputed version of the ordinal covariate under MAR, denoted \( X_{1m}^{\mar} = \{ X_1^{\obs}, X_{1m}^{\mar} \} \), where \( X_{1m}^{\mar} \) denotes the \( m \)-th imputed version of the missing values in \( X_1 \), for \( m = 1, \dots, M \).

\noindent \textbf{2. Threshold-based MNAR adjustment.} \\
The goal of this step is to transform the MAR-based imputations \( X_{1m}^{\mar} \) into adjusted values \( X_{1m}^{\mnar} \), under user-specified MNAR assumptions. This is achieved by perturbing the latent thresholds used to define ordinal categories.

\vspace{0.5cm}

\begin{algorithm}[H]
\small
\caption{Delta-adjustment algorithm for ordinal covariates}
\begin{algorithmic}
\STATE Repeat for each \( m = 1, 2, \dots, M \). \hfill \COMMENT{\( M \): number of imputations}
\STATE Fit ordinal regression \( X_{1m}^{\mar} \sim f(Y, \mat{X}_{\{-1\}}) \); obtain \( \widehat{\beta}_m, \widehat{\zeta}_m \). \hfill \COMMENT{ Coefficients and thresholds}
\STATE Compute \( \eta_{im} = [Y_i, \mat{X}_{i,\{-1\}}]^\top \widehat{\beta}_m \) for all \( i \) with \( R_i = 0 \). \hfill \COMMENT{Expected value of latent variable}
\STATE Generate \( \widehat{\theta}_{im}^* = \eta_{im} + \varepsilon_i \), with \( \varepsilon_i \sim \mathcal{N}(0, \sigma^2) \). \hfill \COMMENT{Stochastic perturbation}
\STATE Shift thresholds: \( \widehat{\zeta}_{km}^* = \widehat{\zeta}_{km} + \delta_k \), for \( k = 1, \dots, K-1 \). \hfill \COMMENT{Category-specific MNAR adjustment}
\STATE Assign \( X_{1im}^{\mnar} = k \) if \( \widehat{\theta}_{im}^* \in (\widehat{\zeta}_{k-1,m}^*, \widehat{\zeta}_{k,m}^*] \), for each \( i \) with \( R_i = 0 \). \hfill \COMMENT{Reclassification}
\STATE Form completed variable \( X_{1m}^{\mnar} = \{ X_1^{\obs}, X_1^{\mnar} \} \). \hfill \COMMENT{Final MNAR-adjusted dataset}
\end{algorithmic}
\end{algorithm}
The interpretation and selection of the sensitivity parameter vector \( \delta \) are critical. We summarize below key practical considerations for guiding its specification.

\begin{diagnostic}
\label{miss_diagnostic}
\end{diagnostic}

(i) The vector \( \delta = (\delta_1, \dots, \delta_{K-1}) \) is external to the data and has the same length as the number of cutpoints (i.e., $K-1$).

(ii) \( \delta \) must reflect plausible modifications in category proportions under MNAR, based on substantive knowledge or prior data. Specifically, across imputations, the average category distribution for \( X_1^{\mnar} \) among missing cases should better align with expert-informed expectations.  Our algorithm is flexible and \( \delta \) may vary by stratum if justified.

 \textbf{3. Analyze the modified data under MNAR.} \\
Each of the \( M \) completed datasets \( \{ X_{1m}^{\mnar}, Y, \mat{X}_{\{-1\}} \} \) is analyzed using the prespecified outcome model linking \( Y \) to the covariates. The resulting estimates and their variances are then combined using Rubin’s rules \citep{carpenter2023, Little2020, buuren2011, grund2023}, which apply regardless of the imputation model used in steps 1 and 2.

\subsection{Method justification}\label{method:justify:chp3}

To justify our method, we rely on a graphical and probabilistic perspective. The DAGs in Figure~\ref{dag-MNAR} illustrate that under an MNAR mechanism, the path \( R \leftarrow X_1 \) remains open. Thus, multiple imputation under MAR fails to block this path, potentially leading to biased inference in the outcome model.

To model the ordinal covariate \( X_1 \), we use a cumulative probit model grounded in the latent variable formulation. Specifically, for each imputed dataset \( m = 1, \dots, M \), we estimate:
\begin{equation*}
    \widehat{\theta}_{im}^* = \widehat{\beta}^y_{m}Y_i + \widehat{\beta}^x_{m} \mat{X}_{i,\{-1\}} + \varepsilon_i,
\end{equation*}
where \( \varepsilon_i \sim \mathcal{N}(0, \sigma^2) \) is an error term introduced to maintain stochasticity in the adjustment process.

Under the MAR assumption, the probability of observing category \( k \) for unit \( i \) is modeled as:
\begin{equation*}
    P(X_{1im}^{\mar} = k \mid Y_i, \mat{X}_{i,\{-1\}}) = \Phi( \widehat{\zeta}^{k}_m - \eta_{im}) - \Phi( \widehat{\zeta}^{k-1}_m - \eta_{im}),
\end{equation*}
where \( \eta_{im} = \widehat{\beta}^y_m Y_i + \widehat{\beta}^x_m \mat{X}_{i,\{-1\}} \), and \(\Phi(\cdot)\) denotes the standard normal cumulative distribution function.

To explore departures from MAR, we shift the threshold parameters using a user-specified sensitivity vector \( \boldsymbol{\delta} = (\delta_1, \dots, \delta_{K-1}) \), resulting in adjusted cutpoints:
\[
\widehat{\zeta}_{km}^* = \widehat{\zeta}_{km} + \delta_k.
\]
Importantly, the regression coefficients \( \widehat{\beta}^y_m, \widehat{\beta}^x_m \) remain invariant under this transformation \citep{agresti2010}, which enables selective alteration of the latent scale classification without affecting the underlying linear predictor. In this way, we generate plausible MNAR scenarios while preserving interpretability.

\paragraph{Stochastic perturbation on the latent scale.}
To prevent the adjusted imputations from becoming deterministic transformations of the MAR imputations, we introduce a Gaussian perturbation \( \varepsilon_i \sim \mathcal{N}(0, \sigma^2) \) with \(\sigma^2 = 1.2\). While the classical probit model assumes unit variance for identification, we relax this constraint in the MNAR adjustment phase to allow additional variability. This added dispersion reflects latent uncertainty under the MNAR mechanism and helps ensure that the modified values preserve stochasticity across imputations, which is a core requirement of Rubin-consistent multiple imputation.

\paragraph{Interpretability of the sensitivity parameter.}
The vector \( \boldsymbol{\delta} \) perturbs the latent cutpoints and thus modifies the probability distribution of imputed categories. Its direction is not inherently conservative or anti-conservative. Therefore, to fully assess the impact of each MNAR scenario, we report both the point estimates and their relative biases with respect to the true values.

\paragraph{Applicability.}
This method is compatible with both non-hierarchical (\texttt{mice}) and hierarchical (\texttt{jomo}) data structures. It offers a flexible framework to examine the robustness of inference on the regression coefficients of interest across a range of user-specified MNAR assumptions. The next section presents a simulation study that evaluates the performance of the proposed approach.

\section{Simulation study} \label{simulation:study:chap3}

We designed a simulation study under both non-hierarchical and hierarchical settings, guided by a model-based data generation strategy. In each case, the partially observed covariate \( X_1 \) is ordinal and subject to a monotone MNAR mechanism. The main scenario focuses on missingness in extreme categories of \( X_1 \); additional results  are provided in Supplementary Section \ref{sup:material}.

\subsection*{(i) Non-hierarchical context (missing extreme categories)}

We simulated data under a generalized linear model in which the probability of the binary outcome is given by \( P(Y = 1 \mid \mat{X}) = \exp(\mat{X}^\top \beta) / (1 + \exp(\mat{X}^\top \beta)) \), where \( \mat{X} = (X_1, X_2) \). The ordinal covariate \( X_1 \in \{1, 2, 3, 4, 5\} \) was generated using a multinomial logistic model conditional on a nominal covariate \( X_2 \in \{1, 2, 3, 4\} \), then transformed into ordered categories. To reflect the characteristics of our real dataset, extreme categories (\( X_1 = 1 \) and \( X_1 = 5 \)) were deliberately overrepresented. The true parameter values are: \( \beta_0 = -1.5 \) for the intercept; \( \beta_{12} = 1 \), \( \beta_{13} = -2 \), \( \beta_{14} = 1.5 \), and \( \beta_{15} = 2 \) for \( X_1 \) (with category 1 as reference); and \( \beta_{22} = 2 \), \( \beta_{23} = 1 \), \( \beta_{24} = 2 \) for \( X_2 \) (with category 1 as reference). We generated 1 000 independent datasets of size \( n = 2000 \). The MNAR mechanism was as follows: if \( Y = 1 \), then 30\% of the values in category \( X_1 = 1 \) were set to missing; if \( Y = 0 \), then 30\% of the values in category \( X_1 = 5 \) were set to missing.

\subsection*{(ii) Hierarchical context (missing extreme categories)}

For the hierarchical scenario, we simulated data from a generalized linear mixed-effects model where the probability of success is defined by \( P(Y = 1 \mid \mat{X}, \mat{Z}) = \exp(\mat{X}^\top \beta + \mat{Z}^\top u) / (1 + \exp(\mat{X}^\top \beta + \mat{Z}^\top u)) \). Here, \( \mat{X} = (X_1, X_2) \), and \( \mat{Z} \) is the design matrix for the random effects. The random effects \( u \) were generated from a normal distribution with mean 0 and standard deviation 0.45. The ordinal covariate \( X_1 \) had \( K = 3 \) categories to match the structure of the real data. The covariate \( X_2 \) was simulated identically to the non-hierarchical case. We assumed a random intercept model. The true parameters were: \( \beta_0 = -1 \) for the intercept; \( \beta_{12} = 1 \) and \( \beta_{13} = -2 \) for \( X_1 \); and \( \beta_{22} = 2 \), \( \beta_{23} = 1 \), and \( \beta_{24} = 2 \) for \( X_2 \).
We generated 500 datasets of \( n = 2000 \) observations each, grouped into 10 clusters of 200 units. Missingness in \( X_1 \) followed an MNAR mechanism conditional on the value of \( Y \) and the stratum defined by \( X_2 \):

 Stratum \( X_2 = 1 \): if \( Y = 1 \), 20\% of values in category \( X_1 = 1 \) were set to missing; if \( Y = 0 \), 30\% of values in \( X_1 = 3 \) were missing.\\
Stratum \( X_2 = 2 \): if \( Y = 1 \), 10\% of values in \( X_1 = 1 \) were missing; if \( Y = 0 \), 40\% of values in \( X_1 = 3 \) were missing.\\
Stratum \( X_2 = 3 \): if \( Y = 1 \), 40\% of values in \( X_1 = 1 \) were missing; if \( Y = 0 \), 10\% of values in \( X_1 = 3 \) were missing.\\
Stratum \( X_2 = 4 \): if \( Y = 1 \), 10\% of values in \( X_1 = 1 \) were missing; if \( Y = 0 \), 30\% of values in \( X_1 = 3 \) were missing.

All simulations and analyses were conducted in \texttt{R} version 4.3.0. The next section presents the simulation results.

\section{Simulation results}\label{simulation:result:chp3}
 
This section presents the results and interpretations of the non-hierarchical context \ref{resul:non:hierachical:ext} and hierarchical context \ref{resul:hierachical}. 
The case where missing values occur on the intermediate categories 
is presented in the supplementary Material Section \ref{non:hierar:intermediaire}.

\begin{figure}[H]  
    \centering
    \includegraphics[width=0.95\textwidth]{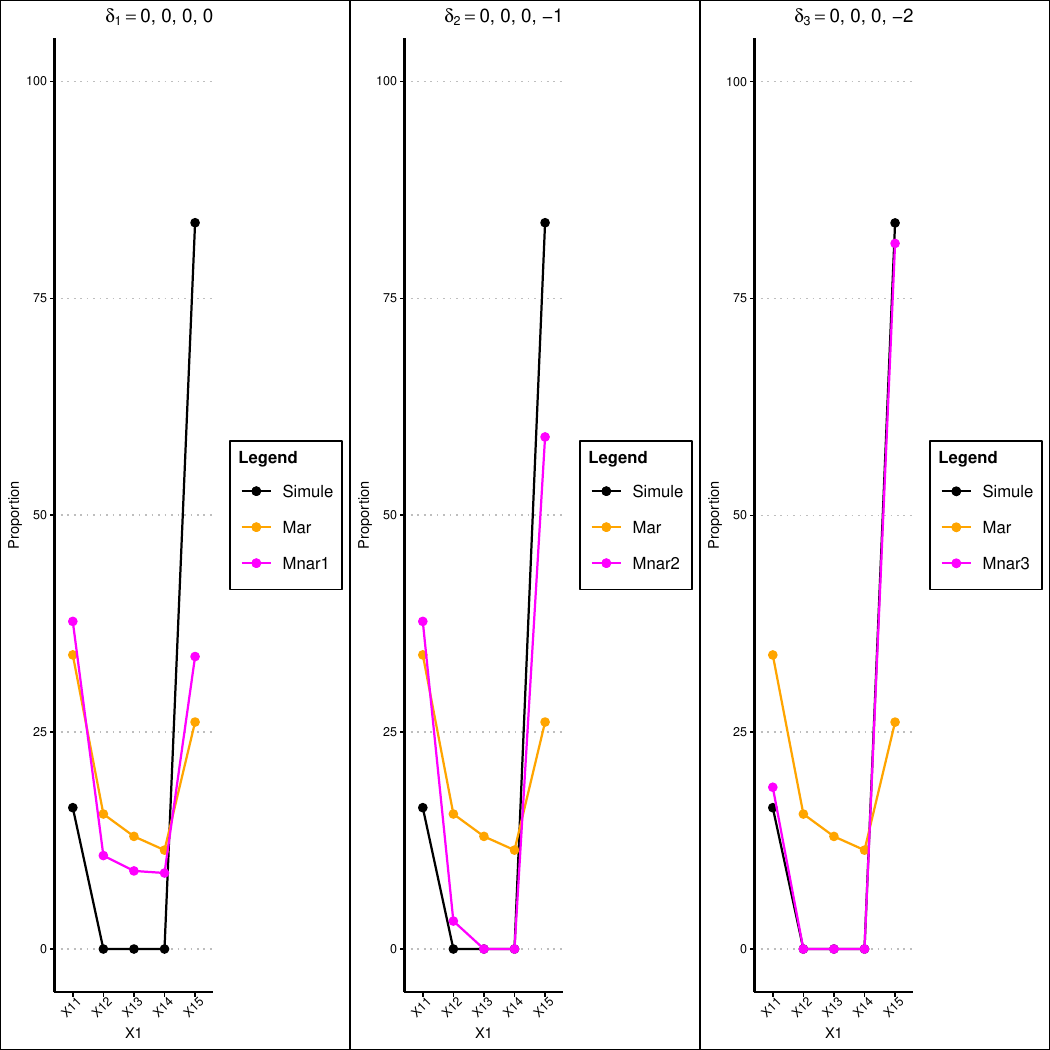}
    \caption{Non-hierarchical context (missing extreme categories). Proportions of the ordinal covariate \( X_1 \) among cases with missing values (\( R = 0 \)), averaged across 1 000 simulations and 10 imputations using \texttt{mice}: Simulated (complete data), MAR, and MNAR scenarios with sensitivity vectors \( \delta_1 = (0,0,0,0) \), \( \delta_2 = (0,0,0,-1) \), and \( \delta_3 = (0,0,0,-2) \).}
    \label{delta_no_hiercachic__ext}
\end{figure}

\subsection{Non-hierarchical context results (missing extreme categories)} \label{resul:non:hierachical:ext}

\ba. \textbf{Selection of the sensitivity parameter vector.}  
As the ordinal covariate \( X_1 \) comprises \( K = 5 \) categories, the sensitivity parameter vector \( \delta \) has dimension \( K - 1 = 4 \). In Figure~\ref{delta_no_hiercachic__ext}, we examine three MNAR scenarios for the imputed values of \( X_1 \): Mnar1, Mnar2, and Mnar3, corresponding respectively to \( \delta_1 = (0, 0, 0, 0) \), \( \delta_2 = (0, 0, 0, -1) \), and \( \delta_3 = (0, 0, 0, -2) \). These vectors reflect increasing deviations from the MAR assumption.

Using the average simulated proportions as a benchmark, we observe that the distributions under MAR and Mnar1 are nearly indistinguishable, despite Mnar1 being nominally defined as an MNAR scenario. This indicates that \( \delta_1 \) does not induce a meaningful departure from MAR (see Section~\ref{Discussion:chap3}). In contrast, the distributions under Mnar2 and Mnar3 more closely resemble the structure observed in the complete (non-missing) data.

\begin{itemize}
    \item[(i)] Both vectors \( \delta_2 = (0, 0, 0, -1) \) and \( \delta_3 = (0, 0, 0, -2) \) tend to reduce the average proportions of intermediate categories (\( X_{12} = 2 \), \( X_{13} = 3 \), \( X_{14} = 4 \)) while increasing the proportions of extreme categories (\( X_{11} = 1 \), \( X_{15} = 5 \)).
    \item[(ii)] The proportion of category \( X_{15} = 5 \) remains higher than that of \( X_{11} = 1 \), consistent with the structure of the simulated ordinal variable \( X_1 \).
\end{itemize}

These observations validate the relevance of our diagnostic criteria (Diagnostic~\ref{miss_diagnostic}) for selecting plausible sensitivity parameters. Notably, Mnar3 yields a category distribution nearly identical to that of the complete data.

Based on this reasoning, we define the following MNAR-adjusted versions:

\begin{align*}
    \text{MNAR}_1 &= \begin{cases} 
        \text{Mnar1} & \text{if $X_1$ is missing} \\
        X_1 \; \text{(Observed)} & \text{if $X_1$ is not missing}
    \end{cases} \\[10pt]
    \text{MNAR}_2 &= \begin{cases} 
        \text{Mnar2} & \text{if $X_1$ is missing} \\
         X_1 \; \text{(Observed)} & \text{if $X_1$ is not missing}
    \end{cases} \\[10pt]
   \text{MNAR}_3 &= \begin{cases} 
        \text{Mnar3} & \text{if $X_1$ is missing} \\
         X_1 \; \text{(Observed)} & \text{if $X_1$ is not missing}
    \end{cases}
\end{align*}

In real-data contexts, expert input remains essential when selecting \( \delta \), especially when the true structure is unknown. Moreover, as discussed later (Section~\ref{Discussion:chap3}), different sensitivity parameter vectors may yield similar results.

\bb. \textbf{Relative bias, empirical standard deviation, and coverage rate.}  
Table~\ref{tab_no_hiercachic__ext} summarizes the performance of the different approaches.

\begin{itemize}
    \item Under MAR and \( \text{MNAR}_1 \), the relative bias is particularly large, especially for coefficients associated with intermediate categories of \( X_1 \), and the coverage rates fall below the nominal level.
    \item The complete-case (CC) analysis yields reduced bias compared to MAR and \( \text{MNAR}_1 \), but its estimates are more variable and coverage is generally unsatisfactory.
    \item In contrast, \( \text{MNAR}_2 \) and \( \text{MNAR}_3 \) produce estimates that are less biased than MAR and more precise than CC, with better coverage.
\end{itemize}

These findings illustrate that well-calibrated MNAR scenarios—guided by appropriate sensitivity vectors—can outperform both MAR-based imputation and complete-case analysis.

\vspace{0.5cm}

\begin{table}[H]
 \caption{Non-hierarchical context (missing extreme categories). Relative bias, empirical standard deviation, and $95\%$ confidence interval coverage rate: SIMULATED (Full data without missing values); CC; MAR; MNAR ($\text{MNAR}_1$, $\text{MNAR}_2$ and $\text{MNAR}_3$)}
  \centering
\begin{adjustbox}{width=0.95\textwidth}
\small
 \begin{tabular}{lcccccccccr}
 \rowcolor{gray!40}   & \textbf{SIMULATED} & \textbf{CC} & \textbf{MAR} & $\textbf{MNAR}_1$ & $\textbf{MNAR}_2$ & $\textbf{MNAR}_3$  \\
 \hline
 \textbf{Relative bias} ($\%$)    &&&&&&\\
    intercept &  0.61  & 0.61 &   0.48 &   0.59 &  2.41 &  3.70 \\
     $X_{12}$  & -0.20 &45.06 & 46.12 & 35.00& 28.78 &26.03\\
      $X_{13}$ &  0.33  & 0.33&  -11.43  & -8.34 &  3.35  & 6.46\\
      $X_{14}$ & 0.33 &15.44 &  8.88 &  8.17&  8.67  &9.15\\
     $X_{15}$ &1.06 &  1.09  & -9.94&   -6.72 &  0.88 & -0.23 \\
   $X_{22}$ &  0.70 & 0.80 & -6.93  &-4.49 & 0.64&  0.96\\
   $X_{23}$ & 0.11  &0.02 &-17.47& -11.61 &-1.18& -0.52 \\
   $X_{24}$ &0.82&  0.79 & -6.85 & -4.42 & 0.72 & 1.06\\
\rowcolor{gray!20} \textbf{Empirical standard deviation}    &&&&&&\\
\rowcolor{gray!20} intercept &   0.14& 0.14 &0.13 & 0.13 & 0.14  &0.14\\
\rowcolor{gray!20}     $X_{12}$  & 0.12& 0.12& 0.12 & 0.12 & 0.12&  0.12\\
 \rowcolor{gray!20}     $X_{13}$ &  0.16 & 0.16 & 0.14&   0.14 &  0.16&   0.16\\
\rowcolor{gray!20}      $X_{14}$ & 0.15& 0.15 &0.14 & 0.14&  0.15 & 0.15\\
\rowcolor{gray!20}     $X_{15}$ & 0.18 &0.18 &0.15&  0.16  &0.18&  0.18 \\
\rowcolor{gray!20}   $X_{22}$ & 0.18 &0.19 &0.17 & 0.17  &0.18 & 0.18\\
\rowcolor{gray!20}   $X_{23}$ & 0.18 &0.19 &0.17 & 0.17  &0.18 & 0.18\\
\rowcolor{gray!20}   $X_{24}$ & 0.18 &0.19 &0.17 & 0.17  &0.18 & 0.18\\
\textbf{Coverage rate}    &&&&&\\
intercept &   0.94 &0.93 &0.95 & 0.94 & 0.93 & 0.93\\
     $X_{12}$  & 0.95 &0.05& 0.04&  0.20&  0.39 & 0.50\\
      $X_{13}$ &0.94& 0.95 &0.74 & 0.85 & 0.95 & 0.89 \\
      $X_{14}$ & 0.95& 0.66 &0.88 & 0.89 & 0.88 & 0.86\\
     $X_{15}$ &  0.95 &0.95 &0.80  &0.89 & 0.95&  0.95 \\
   $X_{22}$ & 0.95& 0.94& 0.90 & 0.94 & 0.95 & 0.95\\
   $X_{23}$ &  0.96 &0.94& 0.84 & 0.91 & 0.95&  0.95 \\
   $X_{24}$ &0.95 &0.95& 0.88 & 0.93 & 0.95 & 0.95\\
\hline
  \end{tabular}
 \end{adjustbox}
     \label{tab_no_hiercachic__ext}
\end{table}

\subsection{Hierarchical context results (missing extreme categories)}\label{resul:hierachical}

\ba. \textbf{Selection of sensitivity parameter vector.}  
As the ordinal variable \( X_1 \) has \( K = 3 \) categories, the sensitivity vector \( \delta \) is of length \( K - 1 = 2 \). We exclude the case \( \delta = (0, 0) \), which closely resembles MAR (see Section~\ref{Discussion:chap3}).

\begin{figure}[H]
    \centering
    \includegraphics[width=0.95\textwidth]{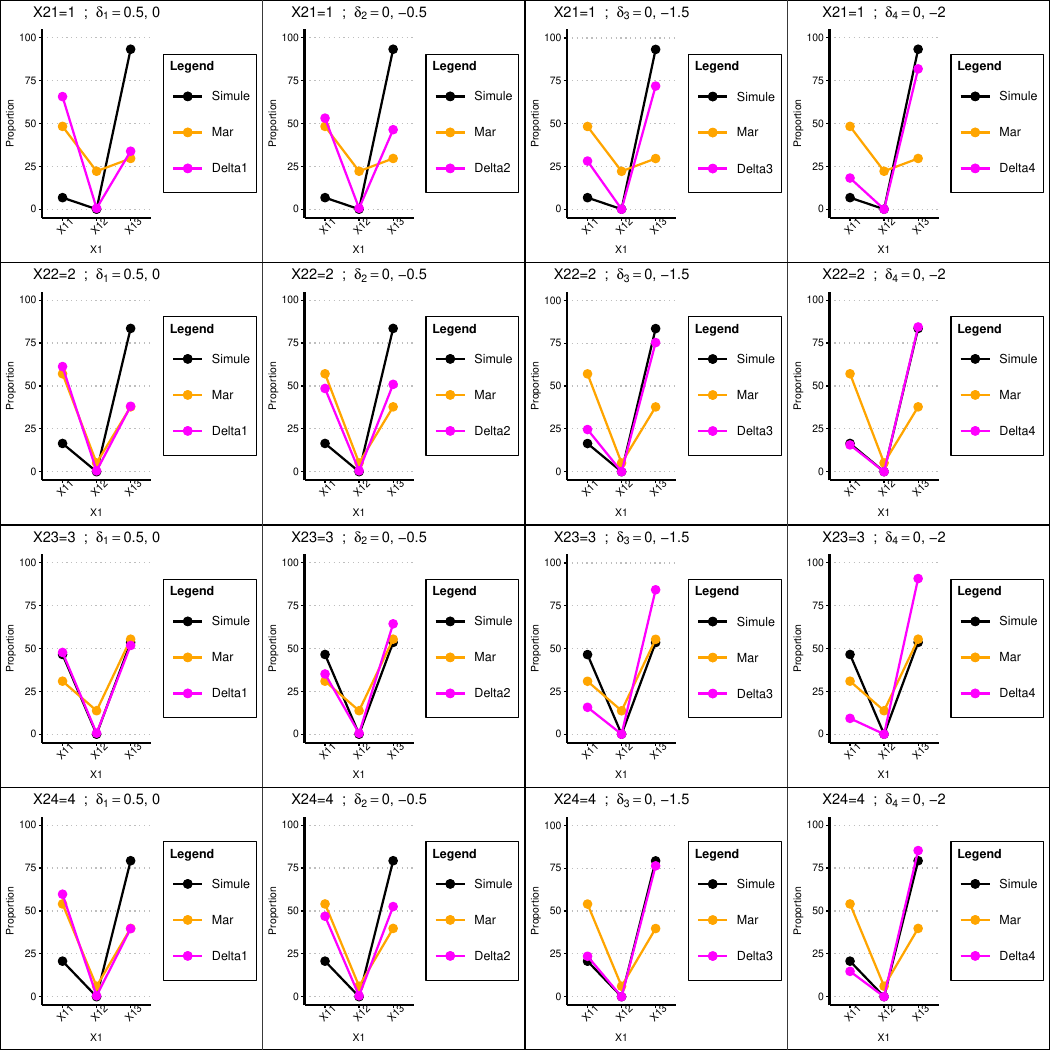}
   \caption{Hierarchical context (missing extreme categories). Proportions of the ordinal variable \( X_1 \) when missing (\( R = 0 \)), averaged over 500 simulations with 10 imputations using the \textit{jomo} package, stratified by \( X_2 \). Delta1–4 correspond respectively to \( \delta_1 = (0.5, 0) \), \( \delta_2 = (0, -0.5) \), \( \delta_3 = (0, -1.5) \), and \( \delta_4 = (0, -2) \).}
   \label{delta_hiercachic__ext}
\end{figure}

(i) For cases with missing values (\( R = 0 \)), our diagnostic (\ref{miss_diagnostic}) guides the identification of sensitivity vectors that substantially modify category distributions under MAR. Four candidate vectors were defined: \( \delta_1 = (0.5, 0) \), \( \delta_2 = (0, -0.5) \), \( \delta_3 = (0, -1.5) \), and \( \delta_4 = (0, -2) \), generating the modified ordinal variables Delta1, Delta2, Delta3, and Delta4. Figure~\ref{delta_hiercachic__ext} illustrates the resulting distributions stratified by \( X_2 \).

(ii) Based on these vectors, we define three MNAR-adjusted versions of \( X_1 \), namely \( \text{MNAR}_1 \), \( \text{MNAR}_2 \), and \( \text{MNAR}_3 \), by selecting, for each stratum of \( X_2 \), the most plausible Delta version according to specific scenarios.

\textbf{Scenario 1.} For each \( X_2 \) stratum, we compare the distributions of Delta1–4 with that of the fully observed data to construct \( \text{MNAR}_1 \). This scenario provides a useful benchmark but cannot be implemented in real data settings since the true distribution is unobserved.

\begin{align*}
\text{MNAR}_1 = \begin{cases} 
\text{Delta4} & \text{if $X_1$ is missing for stratum $X_{21}$} \\
\text{Delta3} & \text{if $X_1$ is missing for stratum $X_{22}$} \\
\text{Delta1} & \text{if $X_1$ is missing for stratum $X_{23}$} \\
\text{Delta3} & \text{if $X_1$ is missing for stratum $X_{24}$} \\
X_1 \text{ (Observed)} & \text{otherwise}
\end{cases}
\end{align*}

\textbf{Scenario 2.} For each stratum, we select the \( \delta \) vector that minimizes the proportion of the intermediate category \( X_{12} = 2 \), while favoring extreme categories \( X_{11} = 1 \) and \( X_{13} = 3 \), with \( X_{13} \) allowed to dominate, consistent with the data-generating structure. This strategy, grounded in our diagnostic, defines \( \text{MNAR}_2 \), applicable in real data settings.

\begin{align*}
\text{MNAR}_2 = \begin{cases} 
\text{Delta2} & \text{if $X_1$ is missing for stratum $X_{21}$} \\
\text{Delta4} & \text{if $X_1$ is missing for stratum $X_{22}$} \\
\text{Delta1} & \text{if $X_1$ is missing for stratum $X_{23}$} \\
\text{Delta4} & \text{if $X_1$ is missing for stratum $X_{24}$} \\
X_1 \text{ (Observed)} & \text{otherwise}
\end{cases}
\end{align*}

\textbf{Scenario 3.} This scenario prioritizes \( X_{13} = 3 \) as the most frequent category across strata, in line with the simulated structure, without strictly minimizing the intermediate category. Here, Delta4 is uniformly applied to all strata, yielding \( \text{MNAR}_3 \). To avoid unrealistic imputations, situations where only \( X_{11} = 1 \) or \( X_{13} = 3 \) dominate should be avoided (see Section~\ref{Discussion:chap3}).

\begin{align*}
\text{MNAR}_3 = \begin{cases} 
\text{Delta4} & \text{if $X_1$ is missing for any $X_2$ stratum} \\
X_1 \text{ (Observed)} & \text{otherwise}
\end{cases}
\end{align*}

\bb. \textbf{Relative biases, empirical standard deviation and coverage rate.}  
Table~\ref{tab_hiercachic} shows that relative biases are substantially higher under MAR, followed by CC, compared to the MNAR-adjusted scenarios. Precision is also improved under \( \text{MNAR}_1 \), \( \text{MNAR}_2 \), and \( \text{MNAR}_3 \), as reflected in smaller empirical standard deviations. Furthermore, MAR and CC yield zero coverage for some estimates, underscoring their unreliability in this context. In contrast, MNAR scenarios consistently provide less biased, more precise estimates with better coverage.

\vspace{0.5cm}

\begin{table}[H]
\caption{Hierarchical context (missing extreme categories). Relative bias, empirical standard deviation, and $95\%$ confidence interval coverage rate of SIMULATED  (Full data without missing values), CC, MAR and MNAR ($\text{MNAR}_1$, $\text{MNAR}_2$ and  $\text{MNAR}_3$)}
  \centering
\begin{adjustbox}{width=0.95\textwidth}
\small
\begin{tabular}{lccccccr}
\rowcolor{gray!40}    & \textbf{SIMULATED} & \textbf{CC} & \textbf{MAR} & $\textbf{MNAR}_1$ & $\textbf{MNAR}_2$ & $\textbf{MNAR}_3$  \\
\hline
\textbf{Relative bias}  ($\%$)    &&&&&&\\
     intercept &  0.49 & -3.69 & 186.86 & 3.80 &-0.55 & 4.12\\
     $X_{12}$  &1.01 & 40.44  &242.61& 21.76& 20.66 &22.36\\
      $X_{13}$ &0.44 & -0.85 &-199.23 & 2.39 & 0.70 & 3.01\\
   $X_{22}$ &1.14  & 8.27 &   5.09&  2.88 &-0.56  &2.49\\
   $X_{23}$ &1.40 &-20.16 &  -8.98& -0.66 &-5.66 &-3.79\\
   $X_{24}$ &0.61 &  3.86  &  3.20 & 1.86& -1.41 & 1.62\\
 
\rowcolor{gray!20}  \textbf{Empirical standard deviation}   &&&&&& \\
\rowcolor{gray!20} intercept & 0.19& 0.19& 0.22 & 0.19 & 0.19 & 0.19\\
 \rowcolor{gray!20}    $X_{12}$  &0.09& 0.09 &0.20 & 0.08 & 0.08 & 0.08\\
 \rowcolor{gray!20}    $X_{13}$ &0.14& 0.14& 0.12&  0.14 & 0.14 & 0.14 \\
 \rowcolor{gray!20}  $X_{22}$ &0.17& 0.17 &0.17 & 0.17 & 0.17 & 0.17\\
\rowcolor{gray!20}   $X_{23}$ &0.17 &0.17 &0.17  &0.17&  0.17 & 0.17\\
\rowcolor{gray!20}   $X_{24}$ &0.16 &0.16& 0.17  &0.16 & 0.16 & 0.16\\
 
  \textbf{Coverage rate}    &&&&&&\\
     intercept &   0.92& 0.92 &0.00 & 0.91 & 0.92  &0.91\\
     $X_{12}$  &0.95& 0.00 &0.00 & 0.34&  0.39 & 0.31\\
      $X_{13}$  &0.96& 0.96 &0.00 & 0.94  &0.96 & 0.94 \\
   $X_{22}$ &0.95& 0.89 &0.93 & 0.95 & 0.95 & 0.95\\
   $X_{23}$ &0.95& 0.81& 0.93 & 0.95&  0.94&  0.94\\
   $X_{24}$ &0.96& 0.95 &0.94 & 0.96  &0.96 & 0.96\\
 \hline
  \end{tabular}
 \end{adjustbox}
    \label{tab_hiercachic}
\end{table}

\section{Application}\label{Application:chap3}

This section presents  the Pan-Canadian trauma data in subsection \ref{trauma:data:desciption}, and the main results  in subsection \ref{trauma:result:chap3}.

\subsection{Pan-Canadian trauma data} \label{trauma:data:desciption}

In Canada, injury is a major public health concern and places a significant burden on society in terms of mortality, morbidity, and costs \citep{TraumaAssociation2011,Haas2011,evans2014,moore2023}. To address this issue, trauma systems have been implemented in many provinces. The objective of the analysis is to compare the mortality risk between provinces.
We used denominalized data on injury hospitalisations from six Canadian provinces (Alberta, British Columbia, Quebec, Nova Scotia, New Brunswick and Ontario). For confidentiality reasons, provinces are coded from A to F. We included all adults ($\geq 16$ years old) hospitalized in trauma centers for major trauma (Injury Severity Score, $\text{ISS} \geq12$) between April 1, 2013 and March 31, 2018. Patients admitted for burns, drowning, poisoning, foreign body ingestion and late sequelae, as well as any deaths on arrival, are excluded from the study. 
This is a retrospective multicenter cohort study of 54354 patients  admitted to trauma centers made up of three levels of care. The variables extracted include the severity of injuries to the head, thorax, abdomen, spine, upper and lower extremities, the patient's sex, the mechanism of injury, and the hospital discharge status (Table \ref{tab.realdata}). Physiological variables such as Glasgow Coma Scale (GCS)  and Systolic Blood Pressure (SBP) are also extracted. 
GCS ranges from the most severe (no response to stimuli, GCS = 3) to the least severe (normal level of consciousness, GCS = 15). 
Indeed, each patient is assessed on the visual, motor, and verbal components, and the GCS is obtained by summing these components \citep{teasdale1976,Borgialli2016}.   
 For clinical interpretation, the GCS is often classified into three  categories: $3 \leq GCS \leq 8$, (GCS1) for severe traumatic brain injury, $9 \leq GCS \leq 12$, (GCS2) for moderate traumatic brain injury, and $13 \leq GCS \leq 15$, (GCS3) for mild traumatic brain injury \citep{Moore2005, moore2023, teasdale2014}.\\

In our data, GCS2 appears to be underrepresented compared to GCS1 and GCS3. Additionally, GCS has  $15.47\%$ missing values, which varies from  province stratum (Table \ref{tab.realdata}). 
The proportions of missing values in each province seem to be higher than those observed in  GCS1 and GCS2 (Table \ref{tab.realdata}).
For all provinces, the proportions of missing values of GCS appear to be higher for patients who have experienced minor trauma (head = 0) and major trauma (head = 5) compared to intermediate categories, with a higher emphasis on minor trauma (head = 0) (Table \ref{tab.realdata2}).
This is coherent with information from content experts whereby the GCS is more likely to be missing in patients with minor extracranial injury and in  the sickest patients who are intubated or sedated on arrival.

{
\renewcommand{\arraystretch}{1} 
\begin{center}
\small
{
\setlength{\tabcolsep}{7pt}
\begin{longtable}{lccccccr}
\caption{Patients Characteristics by Province.} \label{tab.realdata} \\
\rowcolor{gray!40} \textbf{Variables} & \textbf{All Provinces, n} ($\%$) & \textbf{A} & \textbf{B} & \textbf{C} & \textbf{D} & \textbf{E} & \textbf{F}  \\ 
\hline 
\endfirsthead

\multicolumn{8}{c}%
{{\bfseries \tablename\ \thetable{} -- (Continued)}} \\
\hline 
\rowcolor{gray!40} \textbf{Variables} & \textbf{All Provinces, n} ($\%$) & \textbf{A} & \textbf{B} & \textbf{C} & \textbf{D} & \textbf{E} & \textbf{F}  \\ 
\hline 
\endhead
\endlastfoot
\textbf{All patients}  & 54354 (100) & 3.81 & 35.02 & 2.00 &  22.09 &11.64 & 25.44\\
\rowcolor{gray!20} \textbf{Died in hospital}  &&&&&&&\\
\rowcolor{gray!20} Yes  & 5838  (10.74) &  13.33& 11.12& 15.02 & 8.83& 12.04& 10.56\\
\rowcolor{gray!20} No  & 48516 (89.26) &  86.67& 88.88 &84.98 &91.17 &87.96 &89.44 \\

 \textbf{Sex}  &&&&&&&\\
 Female &38092 (70.08) &  71.08& 68.17& 75.48& 72.59& 70.57 &69.74\\
 Male &16262 (29.92)& 28.92& 31.83 &24.52& 27.41 &29.43& 30.26 \\
\rowcolor{gray!20} \textbf{Systolic Blood Pressure}  & &&&&&&\\
\rowcolor{gray!20} $< 90$  & 2065 (3.8) & 3.81  &3.30&  4.70 & 3.86 & 3.89 & 4.32\\
\rowcolor{gray!20} $\geq 90$  & 52289 (96.2)& 96.19 &96.70& 95.30 &96.14 &96.11 &95.68\\

\textbf{Transfer}   &&&&&&&\\
Yes &21255 (39.1) & 49.73& 38.66 &45.16& 37.62 &51.10 &33.44\\
No  &33099 (60.9) &50.27& 61.34 &54.84 &62.38 &48.90& 66.56\\

\rowcolor{gray!20} \textbf{Level}   &&&&&&&\\
\rowcolor{gray!20} Designation I & 32723 (60.20)& 82.91 &54.11 &62.30 &67.47& 80.34 &49.50\\
\rowcolor{gray!20} Designation II &  12220 (22.48) & - &22.22 &37.70& 17.62 &19.66 &30.53\\
\rowcolor{gray!20} Designation III & 9411 (17.31)& 17.09 &23.67&  0.00& 14.92 & - &19.97\\

\textbf{Glasgow Coma Scale}   & &&&&&&\\
$3 \leq GCS \leq 8$  & 4745 (8.73) & 5.70 & 8.79 & 8.20 &9.52& 8.50 &8.55\\
$9 \leq GCS \leq 12$  & 2690 (4.95) &3.43 &5.05 &5.81& 5.51 &4.68 &4.61\\
$13 \leq GCS \leq 15$ &38510 (70.85) &57.03 &72.86 &79.45 &74.15& 70.33 &66.85\\
Missing &\textbf{8409 (15.47)} &\textbf{33.85} & \textbf{13.30} & \textbf{6.54} & \textbf{10.83} & \textbf{ 16.48} & \textbf{19.98}\\

\rowcolor{gray!20} \textbf{Age} &&&&&&&\\
\rowcolor{gray!20} 16 - 54  & 25187 (46.34) &44.28& 37.55 &51.43 &57.12& 47.29& 48.55 \\
\rowcolor{gray!20} 55 - 64 & 8375 (15.41) &14.97& 15.88& 16.68& 15.07 &14.94& 15.24\\
\rowcolor{gray!20} 65 - 74 &  7710 (14.18)&15.60& 15.83& 13.46& 11.14 &14.13 &14.43 \\
\rowcolor{gray!20} 75 - 84 &  7682 (14.13) & 15.98 &17.54& 11.89& 10.28 &14.24 &12.65\\
\rowcolor{gray!20} $\geq$ 85 & 5400 (9.93)&9.17 &13.21&  6.54&  6.39 & 9.40 & 9.13\\

\textbf{Mechanism Of Injury}   &&&&&&&\\
Motor vehicle accident & 19830 (36.48) & 36.75& 32.83& 40.92 &40.91& 39.75& 35.78\\
Fall from its height  & 10764 (19.80) & 19.75&  24.01&  20.65 & 19.12&  12.91&  17.69\\
Fall from more than 1 meter & 13070 (24.05) & 23.37& 26.54& 19.82& 17.34& 29.68 &24.29\\
Penetrating trauma & 1926 (3.54) &2.99 & 2.34 & 3.32 & 5.13 & 3.65 & 3.88\\
Other & 8764 (16.12) &17.14& 14.28 &15.30 &17.49& 14.00 &18.36\\

\rowcolor{gray!20} \textbf{Severity of head injuries}$^{a}$  &&&&&&&\\
\rowcolor{gray!20} 1 &20359 (37.46) &28.34& 34.86 &38.53& 42.73 &33.65& 39.48\\
\rowcolor{gray!20} 2 &4121 (7.58) &11.69  &9.00 & 4.24 & 4.17 & 7.11&  8.45\\
\rowcolor{gray!20} 3 &  3657 (6.73)& 3.72  &7.85 & 4.52&  5.99 & 4.66 & 7.40\\
\rowcolor{gray!20} 4 &6503 (11.96) &10.67& 12.87& 13.18 & 9.52& 14.41& 11.82\\
\rowcolor{gray!20} 5 & 8215 (15.11) &19.27& 12.60 &21.94 &19.66 &14.13 &13.93\\
\rowcolor{gray!20} 6 & 11499 (21.16) & 26.32& 22.83& 17.60& 17.93 &26.03 &18.92\\

 \textbf{Severity of thorax injuries}$^{a}$    & &&&&&&\\
 1  & 27231 (50.10) & 37.57& 51.50 &53.09 &51.33& 48.57 &49.45\\
 2  &  2472 (4.55) & 6.81 & 3.19 & 4.79 & 3.22&  4.93  &7.04\\
 3  &  4656 (8.57) & 12.12  &7.90 & 10.60 & 7.60 &10.18 & 8.89\\
 4  & 14815 (27.26) & 29.94 &30.17& 25.99& 22.63 &27.55 & 26.83\\
 5  &  4304 (7.92) & 11.44 & 6.36 & 4.24 &12.86 & 6.40 & 6.23\\
 6  &  876 ( 1.61) & 2.12 & 0.89 & 1.29 & 2.35 & 2.37 & 1.56\\
 
\rowcolor{gray!20} \textbf{Severity of abdomen injuries}$^{a}$    &&&&&&&\\
\rowcolor{gray!20} 1  & 42693 (78.55) &70.45& 81.45 &81.29& 81.83 &77.08 &73.37\\
\rowcolor{gray!20} 2  & 3461 (6.37) & 8.69 & 4.58 & 4.61&  2.60 & 5.18 &12.44\\
\rowcolor{gray!20} 3  &  3940 (7.25) & 9.85&  7.09 & 7.37&  6.86 & 8.84 & 6.68\\
\rowcolor{gray!20} 4  &  2111 (3.88) & 6.57 & 3.62&  4.61 & 3.69 & 4.63 & 3.61\\
\rowcolor{gray!20} 5  &  1762 (3.24) & 3.81 & 2.62&  1.47 & 4.43  &3.19  &3.15\\
\rowcolor{gray!20} 6  &  387 (0.71) & 0.63 & 0.65 & 0.65 & 0.58 & 1.07 & 0.76\\

 \textbf{Severity of spine injuries}$^{a}$    & &&&&&&\\
 1  &35566 (65.43) & 57.41& 65.80 &66.73& 69.18& 63.39 &63.71\\
 2  & 749 (1.38) & 2.61 & 0.38&  1.38 & 1.24 & 2.10  &2.35\\
 3  &10075 (18.54) & 25.59& 18.14& 16.77& 16.87& 20.94 &18.51\\
 4  & 5045 (9.28) & 8.64 &11.10 & 7.00 & 8.50 & 8.01 & 8.32\\
 5  & 1931 (3.55) & 3.33 &  3.53  & 3.13 &  2.72 &  3.15 &  4.56\\
 6  & 988 (1.82) & 2.41 & 1.05&  4.98&  1.49 & 2.40 & 2.55\\

\rowcolor{gray!20}\textbf{Upper extremities injuries}$^{b}$    & &&&&&&\\
\rowcolor{gray!20} 1  &32275 (59.38) & 49.54& 60.09 &65.62 &70.56 &60.60& 49.12\\
\rowcolor{gray!20} 2  & 7472 (13.75) & 14.97 &12.85 &10.23 & 5.10 & 8.33 &25.06\\
\rowcolor{gray!20} 3  & 12971 (23.86) & 32.25 & 25.45 &21.84 &18.66 &28.32 &23.06\\
\rowcolor{gray!20} 4  & 1426 (2.62) & 2.37 & 1.50 & 1.84 & 4.86  &2.40  &2.43\\
\rowcolor{gray!20} 5  & 210 (0.39) & 0.87 & 0.11&  0.46 & 0.83 & 0.35 & 0.33\\
 
 \textbf{Lower extremities injuries}$^{b}$    & &&&&&&\\
 1  &33775 (62.14) & 52.73 &61.05 &67.83 &75.10 &63.73 &52.61\\
 2  & 6644 (12.22) & 13.38 &12.17 & 9.59 & 3.50 & 7.37 &22.13\\
 3  &7051 (12.97) & 16.51& 13.65& 12.26& 10.25& 15.82 &12.62\\
 4  &5586 (10.28) & 14.73 &11.10 & 6.36 & 9.74 &11.06&  8.89\\
 5  & 1298 (2.39) & 2.66  &2.02 & 3.96 & 1.41 & 2.02 & 3.75\\
\hline
&&&&&&&\\
\multicolumn{7}{l}{$^{a}$ Score from 1 to 6 with increasing severity}\\
\multicolumn{7}{l}{$^{b}$ Score from 1 to 5 with increasing severity}\\
\end{longtable}
}
\end{center}
}

{
\renewcommand{\arraystretch}{0.8} 
\begin{center}
\tiny
{
\setlength{\tabcolsep}{12pt}
\begin{table}[H]
    \caption{Proportion of missing GCS by head trauma and transfer status within province strata}
        \begin{tabular}{lccccccccr}
            \rowcolor{gray!40}
            \multicolumn{1}{l}{} & \multicolumn{6}{c}{\textbf{Severity of head injuries}} & \multicolumn{3}{c}{\textbf{Transfer}} \\
            \cmidrule(lr){2-7} \cmidrule(l){9-10}
          \rowcolor{gray!40}  \textbf{Provinces} & 1 & 2 & 3 & 4 & 5 & 6 && No & Yes\\
            A & 19.97 & 6.70 & 4.00 & 8.70 & 23.68 & 36.95 && 48.22 & 51.78 \\
          \rowcolor{gray!20}  B & 62.90 & 6.16 & 3.79 & 8.26 & 7.55 & 11.34 && 65.98 & 34.02 \\
            C & 53.52 & 1.41 & 0.00 & 8.45 & 15.49 & 21.13 && 56.34 & 43.66 \\
          \rowcolor{gray!20}  D & 25.00 & 2.15 & 3.08 & 11.31 & 24.38 & 34.08 && 48.23 & 51.77 \\
            E & 22.72 & 4.60 & 2.40 & 14.00 & 18.41 & 37.87 && 30.49 & 69.51 \\
          \rowcolor{gray!20}  F & 30.66 & 4.96 & 4.85 & 9.84 & 16.18 & 33.51 && 51.97 & 48.03
        \end{tabular}
    \label{tab.realdata2}
\end{table}
}
\end{center}
}
\subsection{Results of Pan-Canadian trauma data analysis}\label{trauma:result:chap3}
 
In Table \ref{tab:delta_table}, Delta1, Delta2, Delta3, and Delta4 represent varying degrees of deviation of MAR, derived respectively from  $\delta_1 = (0.5, 0.5)$, $\delta_2 = (0.3, 0)$, $\delta_3 = (0, -0.3)$, and $\delta_4 = (-0.5, -0.5)$. We refer to the diagnostic (\ref{miss_diagnostic}) approach and expert opinions to select the sensitivity parameter vector. 

\ba. \textbf{Selection of the  sensitivity parameter vector:}
According to the diagnostic (\ref{miss_diagnostic}), for individuals with missing GCS, the parameter vectors $\delta_2 = (0.3, 0)$ and $\delta_3 = (0, -0.3)$ satisfy the following conditions for each  province stratum: (i) the proportions of GCS1 and GCS3 modified  under MNAR are  higher compared to those under MAR and (ii) under MNAR, the proportions of GCS3  are higher compared to those of GCS1. However, $\delta_1 = (0.5, 0.5)$ and $\delta_4 = (-0.5, -0.5)$ do not satisfy both points.
From the above, it is possible to construct:
\begin{align*}
   \text{MNAR}_1 &= \begin{cases} 
        \text{Delta2} & \text{if GCS is missing for all Province strata} \\
        \text{GCS} \;  \text{(Observed)} & \text{if GCS is not missing}
    \end{cases} \\[10pt]  
    \text{MNAR}_2 &= \begin{cases} 
        \text{Delta3} & \text{if GCS is missing for all Province strata} \\
     \text{GCS} \;  \text{(Observed)} & \text{if GCS is not missing}
    \end{cases} \\[10pt]
    \text{MNAR}_3 &= \begin{cases} 
        \text{Delta2} & \text{if GCS is missing for Province A, C, and D} \\
        \text{Delta3} & \text{if GCS is missing for Province B, E, and F} \\
       \text{GCS} \;  \text{(Observed)} & \text{if GCS is not missing}
    \end{cases}
\end{align*}

Where Delta2 and Delta3 represent varying degrees of deviation of MAR.
The modified variables $\text{MNAR}_1$, $\text{MNAR}_2$, and $\text{MNAR}_3$ within a province stratum are not the only possible associations. Other sets of sensitivity parameters vectors or alternative associations within each province stratum can be explored.
\vspace{0.5cm}

\begin{table}[H]
\centering
\caption{Proportions of GCS categories for missing cases ($R = 0$) based on 20 imputations using the \textit{jomo} package, stratified by Province. Delta1–4 correspond respectively to $\delta_1 = (0.5, 0.5)$, $\delta_2 = (0.3, 0)$, $\delta_3 = (0, -0.3)$, and $\delta_4 = (-0.5, -0.5)$.}
\label{tab:delta_table}
\begin{adjustbox}{width=\textwidth}
\begin{tabular}{lrrrrr}
\hline
\rowcolor{gray!40}
&  \makecell{\textbf{MAR} \\ -}
& \makecell{\textbf{Delta1} \\ ($\delta_1 = 0.5, 0.5$)}
& \makecell{\textbf{Delta2} \\ ($\delta_2 = 0.3, 0$)}
& \makecell{\textbf{Delta3} \\ ($\delta_3 = 0, -0.3$)}
& \makecell{\textbf{Delta4} \\ ($\delta_4 = -0.5, -0.5$)} \\
\hline
\textbf{Province A} & & & & & \\
GCS1$^a$ & 17.90 & 32.85 & 27.95 & 21.81 & 13.69 \\
GCS2     & 13.54 &  7.57 &  0.18 &  0.16 &  4.73 \\
GCS3     & 68.56 & 59.57 & 71.88 & 78.02 & 81.58 \\
\rowcolor{gray!20} \textbf{Province B} & & & & & \\
\rowcolor{gray!20} GCS1 & 16.00 & 31.77 & 26.98 & 20.76 & 12.76 \\
\rowcolor{gray!20} GCS2 & 11.56 &  7.33 &  0.18 &  0.15 &  4.62 \\
\rowcolor{gray!20} GCS3 & 72.44 & 60.90 & 72.84 & 79.10 & 82.62 \\
\textbf{Province C} & & & & & \\
GCS1 & 12.25 & 30.07 & 25.35 & 19.72 & 10.42 \\
GCS2 &  8.94 &  7.18 &  0.21 &  0.28 &  5.85 \\
GCS3 & 78.80 & 62.75 & 74.44 & 80.00 & 83.73 \\
\rowcolor{gray!20} \textbf{Province D} & & & & & \\
\rowcolor{gray!20} GCS1 & 16.91 & 32.55 & 27.58 & 21.38 & 13.12 \\
\rowcolor{gray!20} GCS2 & 11.72 &  7.30 &  0.17 &  0.12 &  4.61 \\
\rowcolor{gray!20} GCS3 & 71.37 & 60.15 & 72.25 & 78.50 & 82.27 \\
\textbf{Province E} & & & & & \\
GCS1 & 18.45 & 32.83 & 27.88 & 21.77 & 13.32 \\
GCS2 & 12.76 &  7.44 &  0.17 &  0.20 &  4.84 \\
GCS3 & 68.80 & 59.72 & 71.95 & 78.03 & 81.84 \\
\rowcolor{gray!20} \textbf{Province F} & & & & & \\
\rowcolor{gray!20} GCS1 & 17.35 & 33.52 & 28.63 & 22.22 & 13.70 \\
\rowcolor{gray!20} GCS2 & 12.18 &  7.42 &  0.17 &  0.14 &  4.82 \\
\rowcolor{gray!20} GCS3 & 70.47 & 59.06 & 71.20 & 77.64 & 81.48 \\
\hline
\\
\multicolumn{6}{l}{$^a$: $\text{GCS1}$ for ($3 \leq \text{GCS} \leq 8$); $\text{GCS2}$ for  ($9 \leq \text{GCS} \leq 12$) and $\text{GCS3}$ for  ($ 13 \leq \text{GCS} \leq 15 $.)}
\end{tabular}
\end{adjustbox}
\end{table}

\bb. \textbf{Odd ratio (OR), Confidence intervals (CI), P-value:} Except for the province, which is our variable of interest, all other variables are adjustment variables. 
CC results have odds ratios that  appear different, with wider confidence intervals, compared to those from MAR, $\text{MNAR}_1$, $\text{MNAR}_2$, and $\text{MNAR}_3$ scenarios (Table \ref{tab:OR_real_data}). Indeed, the CC hypothesis seems less plausible, as it is unlikely that these missing values occur by accident, given that their proportion by province stratum appears higher than the observed proportions of GCS1 and GCS2 (Table \ref{tab.realdata}). 

 Under MNAR, using GCS1 as the reference category,  the GCS estimates show significant variations $(\text{P-value }< 0.05)$  in the probabilities of death, compared to  MAR and CC estimates (Table \ref{tab:OR_real_data}). This is a tipping point approach. However, under MAR and all MNAR scenarios, the odds ratio of provinces seems similar. 
Note that even though the province coefficient estimates under MAR are robust,  their interpretation differs depending on whether the treatment of missing values is performed under MAR vs. MNAR. In our real data, MNAR is more plausible than MAR because it accounts for the information not available in the data. Additionally, given the importance of the GCS in trauma data, it would be absurd if these estimates were not significant.

Considering province B as the reference (with the highest proportion of patients), and assuming that the missing values of GCS are specific to patients with GCS1 and GCS3. In province D, being admitted reduces the odds of death by $33\%$, $(\text{P-value} < 0.05)$ in all MNAR scenarios compared to province B. In province F,  $10\%$, $(\text{P-value} > 0.05)$ of reduction is also observed. However, for provinces A, E, and C, the probability of death increases respectively by $32\%$, $(\text{P-value} < 0.05)$, $2\%$, $(\text{P-value} > 0.05)$, and $77\%$, $(\text{P-value} < 0.05)$.

{
\renewcommand{\arraystretch}{1} 
\begin{center}
{
\setlength{\tabcolsep}{11pt}
\begin{table}[H]
\caption{Comparison of mortality Odds Ratios (OR), Confidence Intervals (CI), and P-value under CC, MAR, and  MNAR ($\text{MNAR}_1$, $\text{MNAR}_2$ and $\text{MNAR}_3$). GCS1 for ($ 3 \leq \text{GCS} \leq 8 $) is the reference category; GCS2 for  ($9 \leq \text{GCS} \leq 12$) and GCS3 for  ($ 13 \leq \text{GCS} \leq 15 $). Province B is the reference category.}
\label{tab:OR_real_data}
\scriptsize
\begin{tabular}{lcccccccc}
\rowcolor{gray!40}
\multirow{2}{*}{\textbf{Method}} & \multicolumn{2}{c}{\textbf{Glasgow}} & \multicolumn{5}{c}{\textbf{Province}} \\
\cmidrule(lr){2-3} \cmidrule(lr){4-8}
\rowcolor{gray!40}
& GCS2 & GCS3 & A & C & D & E & F \\
\midrule
\rowcolor{gray!20} \textbf{CC} &&&&&&&\\
OR     & 0.96 & 0.98 & 1.26 & 2.11 & 0.61 & 0.89 & 0.62 \\
CI     & (0.88, 1.04) & (0.86, 1.11) & (0.82, 1.94) & (1.27, 3.53) & (0.45, 0.83) & (0.59, 1.32) & (0.48, 0.81) \\
P-value& 0.26 & 0.73 & 0.29 & 0.00 & 0.00 & 0.55 & 0.00 \\
\rowcolor{gray!20} \textbf{MAR} &&&&&&&\\
OR     & 1.09 & 0.79 & 1.33 & 1.78 & 0.67 & 1.03 & 0.90 \\
CI     & (0.87, 1.36) & (0.59, 1.06) & (0.94, 1.90) & (1.13, 2.80) & (0.51, 0.88) & (0.73, 1.45) & (0.72, 1.13) \\
P-value& 0.43 & 0.11 & 0.11 & 0.01 & 0.00 & 0.88 & 0.37 \\
\rowcolor{gray!20} \textbf{MNAR$_1$} &&&&&&&\\
OR     & 0.88 & 1.18 & 1.32 & 1.77 & 0.67 & 1.02 & 0.90 \\
CI     & (0.81, 0.96) & (1.05, 1.34) & (0.93, 1.88) & (1.13, 2.79) & (0.51, 0.87) & (0.72, 1.44) & (0.72, 1.14) \\
P-value& 0.00 & 0.00 & 0.11 & 0.01 & 0.00 & 0.90 & 0.38 \\
\rowcolor{gray!20} \textbf{MNAR$_2$} &&&&&&&\\
OR     & 0.86 & 1.19 & 1.32 & 1.78 & 0.67 & 1.02 & 0.90 \\
CI     & (0.79, 0.93) & (1.06, 1.35) & (0.93, 1.88) & (1.13, 2.80) & (0.51, 0.87) & (0.72, 1.45) & (0.72, 1.13) \\
P-value& 0.00 & 0.00 & 0.11 & 0.01 & 0.00 & 0.89 & 0.37 \\
\rowcolor{gray!20} \textbf{MNAR$_3$} &&&&&&&\\
OR     & 0.87 & 1.19 & 1.32 & 1.77 & 0.67 & 1.02 & 0.90 \\
CI     & (0.81, 0.93) & (1.05, 1.34) & (0.93, 1.88) & (1.13, 2.79) & (0.51, 0.87) & (0.72, 1.44) & (0.72, 1.13) \\
P-value& 0.00 & 0.00 & 0.12 & 0.01 & 0.00 & 0.90 & 0.38 \\
\bottomrule
\end{tabular}
\end{table}
}
\end{center}
}
\vspace{-1cm}
\begin{tablenotes}
\item Note: The intraclass correlation ranges from $2.19\%$ to $2.86\%$ for all method.
\end{tablenotes}

\section{Discussion}\label{Discussion:chap3}

\textbf{Choice of MI packages.}  
The imputation procedure relies on the \texttt{mice} and \texttt{jomo} packages, as they are among the most widely used for multiple imputation under MAR \citep{Robitzsch2011, honaker2011, Robitzsch2023, gelman2011, audigier2023}. However, our sensitivity analysis approach requires two specific conditions: (i) the imputation model must preserve the ordering of the partially observed ordinal variable; if not, we recommend transforming the imputed variable into an ordinal factor before applying the second step of our method; (ii) the model must account for within-cluster correlation when the data are hierarchical.

\textbf{Construction of the sensitivity parameter vector.}  
The method involves generating a grid of sensitivity parameter vectors to perturb the intercepts from the ordinal regression model. We deliberately avoid reordering the shifted thresholds when subdividing the latent normal variable. Although reordering may produce an ordinal variable with the same categories, it does not account for the missing data process. Any vector with at least one non-zero element can alter the category probabilities by favoring over- or under-representation of specific levels. Our approach discards unrealistic cases where one category absorbs all the probability mass, resulting in $100\%$ representation among imputed values. Furthermore, different vectors can yield nearly identical marginal distributions. To assess plausibility, we compare the average proportions of each category (for $R = 0$) under MAR to those obtained from the perturbed models using our diagnostic tool. Unsurprisingly, the zero vector leads to nearly identical distributions under MNAR and MAR, since the modified intercepts and latent variables remain unchanged.

\textbf{Strengths.}  
Unlike conventional approaches that compare CC and MAR to gauge robustness, our method explores departures from MAR toward plausible MNAR scenarios through direct modification of the imputation model. This framework provides interpretable diagnostics to assess sensitivity of results to violations of the MAR assumption. In our real dataset, most variables—except GCS—showed robustness under MAR, likely due to the large sample size and high proportion of categorical variables. These findings contrast with our simulation studies, which relied on well-balanced synthetic data.

\textbf{Limitations.}  
Our simulations and methods focus on scenarios where missing values affect only two specific categories of an ordinal covariate, mimicking our real dataset. We did not explore more complex missing data patterns, such as cases with more than two missing categories or non-monotone structures. Moreover, the simulated datasets were not derived from the empirical distribution of the real data but rather constructed under idealized assumptions (e.g., balanced design, predefined regression parameters). This gap may limit the direct applicability of simulation-based findings to real-world settings.

Another limitation lies in the specification and interpretation of the sensitivity parameter vector \( \delta \). While our diagnostic aids in identifying plausible vectors, the choice of \( \delta \) remains partially subjective, as it cannot be identified from the data alone and must be informed by expert knowledge. In addition, the variance parameter \( \sigma^2 \) in the latent model is fixed (e.g., to 1.2 in our implementation) and may influence the effective perturbation induced by \( \delta \), yet we did not evaluate the sensitivity of our results to different values of \( \sigma^2 \).

Although \texttt{jomo} handles random effects during the imputation phase, the ordinal regression model used for threshold adjustment does not explicitly incorporate such random effects. Ideally, the second-step model should mirror the structure of the imputation model, particularly in multilevel settings. Lastly, our method is applied to a simplified setting with a single partially observed ordinal covariate. In practice, datasets often involve multiple variables with missing values and complex joint distributions. Extending our framework to such multivariate and high-dimensional missingness patterns is an important avenue for future work.

\textbf{Computational considerations.}  
Parallel computing was used to reduce computation time \citep{RCoreTeam2024}. All code and replication material are publicly available at:  
\url{https://github.com/abdoulaye-dioni/Sensitivity-Analysis}.

\textbf{Potential extensions.}  
Future developments could generalize the proposed sensitivity analysis framework in several directions. First, it could be extended to ordinal variables with more than two missing categories by modifying a higher-dimensional intercept vector and evaluating diagnostic concordance accordingly. Second, the approach could be adapted to non-monotone missing data patterns, using iterative imputation steps within the joint modeling or FCS framework. Third, the method may be extended to nominal or continuous variables by defining sensitivity parameters that operate on latent structures suitable for those variable types (e.g., latent utilities or residual shifts). Finally, extensions to more complex data structures such as longitudinal or multilevel designs may require additional modeling layers to handle dependency across time or clusters while preserving interpretability of the sensitivity parameters.
\vspace{2cm}

\textbf{Acknowledgement:}  Abdoulaye Dioni, would like to acknowledge the Canadian Francophonie Scholarship Program and the government of Burkina Faso for their financial support during his PhD studies.

\textbf{Conflict of interest statement:} The authors have declared no conflict of interest.

\newpage
\appendix

\section*{\Huge{Sensitivity analysis method in the presence of a  missing not at random ordinal independent variable :Supplementary material} }\label{sup:material}
\pagenumbering{Alph}
\vspace{1cm}
\section{Bias of the imputed data under MAR}

\subsection{Bias in the non-hierarchical context (missing extreme categories)}\label{mice:bias:ext}
The coefficients and intercepts obtained from data imputed using \texttt{mice} and \texttt{Jomo} are potentially biased. Additionally, the bias appears to be greater under \texttt{jomo} than under \texttt{mice} (Tables \ref{tab:biais_non_hierarchique} and \ref{tab:biais_hierarchique}). For generating missing data, we assume that if $Y = 1$, then $30\%$ of $X_{11}$ are missing, and if $Y = 0$, then $30\%$ of $X_{15}$ are missing. (Table \ref{tab:biais_non_hierarchique}) compares the coefficients, intercepts, and potential bias with simulated data  and imputed under \texttt{mice} package.

\begin{table}[H]
    \centering
    \caption{Non-hierarchical context (missing extreme categories), Bias of Coefficients under MAR  using \texttt{R} package \texttt{mice} for 1000 simulations and 10 imputations.}
    \begin{adjustbox}{width=\textwidth}
        \begin{tabular}{lcccccr}
          \rowcolor{gray!40}
            \multicolumn{1}{l}{} & \multicolumn{2}{c}{\textbf{SIMULATED}} & \multicolumn{2}{c}{\textbf{MAR}} & \multicolumn{1}{c}{} \\
            \cmidrule(lr){2-3} \cmidrule(lr){4-5} 
         \rowcolor{gray!40}   & Estimate & sd (Estimate) &  Estimate & sd (Estimate)& Relative bias ($\%$)\\
            \textbf{Coefficients} &  &  &  &  &  \\
            $Y_{1}$ & 0.810 & 0.054 &  1.034 & 0.055 &  27.619 \\
           $X_{22}$ & -0.079 & 0.073 & -0.060 & 0.073 & -24.952 \\
           $X_{23}$ &  0.052 & 0.071 &  0.042 & 0.071 & -18.327 \\
           $X_{24}$ & -0.080 & 0.073 & -0.060 & 0.073 & -24.642 \\
          \rowcolor{gray!20}  \textbf{Intercepts} &  &  &  &  &  \\
          \rowcolor{gray!20}  $1|2$ & -0.258 & 0.053 & -0.228 & 0.053 & -11.702 \\
          \rowcolor{gray!20}  $2|3$ & 0.101 & 0.052 & 0.175 & 0.052 & 73.084 \\
          \rowcolor{gray!20} $3|4$ & 0.440 & 0.052 & 0.547 & 0.053 & 24.341 \\
          \rowcolor{gray!20} $4|5$ & 0.797 & 0.055 & 0.945 & 0.056 & 18.590 
        \end{tabular}
    \end{adjustbox}
    \label{tab:biais_non_hierarchique}
\end{table}

\subsection{Bias in the hierarchical context (missing extreme categories)}\label{jomo:bias:ext}
 We assumed a MNAR mechanism for $X_1$, and we allowed missing data to vary depending on the strata levels of the variable $X_2$, considered as an exposure variable of interest, as in real data:
 
Stratum $X_{21}$: if $Y = 1$, then $20\%$ of $X_{11}$ are missing, if $Y = 0$, then $30\%$ of  $X_{13}$ are missing.
Stratum $X_{22}$: if $Y = 1$, then $10\%$ of $X_{11}$ are missing, if $Y = 0$, then $40\%$ of  $X_{13}$ are missing.
Stratum $X_{23}$: if $Y = 1$, then $40\%$ of  $X_{11}$ are missing, if $Y = 0$, then $10\%$ of  $X_{13}$ are missing.
Stratum $X_{24}$: if $Y = 1$, then $10\%$ of  $X_{11}$ are missing, if $Y = 0$, then $30\%$ of $X_{13}$ are missing. \\

Table \ref{tab:biais_hierarchique} compares the coefficients, intercepts, and potential bias with simulated data and imputed under the \texttt{R} package \texttt{jomo}.

\begin{table}[H]
    \centering
    \caption{Hierarchical context (missing extreme categories), bias of coefficients under MAR with the \texttt{R} package \texttt{jomo} for 500 simulations and 10 imputations.}
    \begin{adjustbox}{width= \textwidth}
       \begin{tabular}{lcccccr}
          \rowcolor{gray!40}
            \multicolumn{1}{l}{} & \multicolumn{2}{c}{\textbf{SIMULATED}} & \multicolumn{2}{c}{\textbf{MAR}} & \multicolumn{1}{c}{} \\
            \cmidrule(lr){2-3} \cmidrule(lr){4-5} 
         \rowcolor{gray!40}   & Estimate & sd (Estimate) &  Estimate & sd (Estimate)& Relative bias ($\%$)\\
            \textbf{Coefficients} &  &  &  &  &  \\
            $Y_{1}$ &  0.576& 0.054& 0.840& 0.055&  45.921& \\
           $X_{22}$ &  0.144& 0.075& 0.088& 0.075& -38.477\\
           $X_{23}$ &  0.253& 0.074& 0.362& 0.074&  43.171\\
           $X_{24}$ &  0.142& 0.075& 0.118& 0.075& -16.764\\ 
          \rowcolor{gray!20}  \textbf{Intercepts} &  &  &  &  &  \\
         \rowcolor{gray!20}   $1|2$ & -0.024& 0.055& 0.155& 0.055& -725.764 \\
          \rowcolor{gray!20}  $2|3$ &  0.451& 0.056& 0.673& 0.056&  49.132
        \end{tabular}
    \end{adjustbox}
    \label{tab:biais_hierarchique}
\end{table}

\section{Non-hierarchical context results for binary response (missing intermediate categories )}\label{non:hierar:intermediaire}

 We simulated a generalized linear model, $\text{P}(Y = 1| \mat{X}) = \exp( \mat{X}^t\beta)/\left(1 + \exp( \mat{X}^t\beta)\right)$ where $Y$ is a binary dependent variable, $ \mat{X}= X_1, X_2$ the set of independent variables and $\beta$ the true parameters. $X_2$ is a nominal variable  with four categories obtained by a random draw from 1 to 4. $X_1 = (X_{11},  X_{12}, X_{13}, X_{14}, X_{15}) $ is an ordinal variable with ($K=5$) categories generated by a multinomial regression from $X_2$, then ordered. Furthermore, $X_1$ is simulated such that the intermediate categories ($X_{12}$ and $X_{14}$) are overrepresented compared to ($X_{11}$, $X_{13}$ and $X_{15}$). The true parameters are as follows: $\beta_0 = -1$ for the intercept; $(\beta_{12}, \beta_{13}, \beta_{14}, \beta_{15}) = (1, -1, -1.5, -2)$ for $X_1$, and $(\beta_{22}, \beta_{23}, \beta_{24}) = (2, 1, 2)$ for $X_2$. For missingness generation, we assume if $Y = 1$ then $40\%$ of $X_{12}$ are missing, and if $Y = 0$ then $30\%$ of $X_{14}$ are missing.

\subsection{Selection of sensitivity parameter vector.}
\begin{figure}[H]
    \centering
    \includegraphics[width=0.95\textwidth]{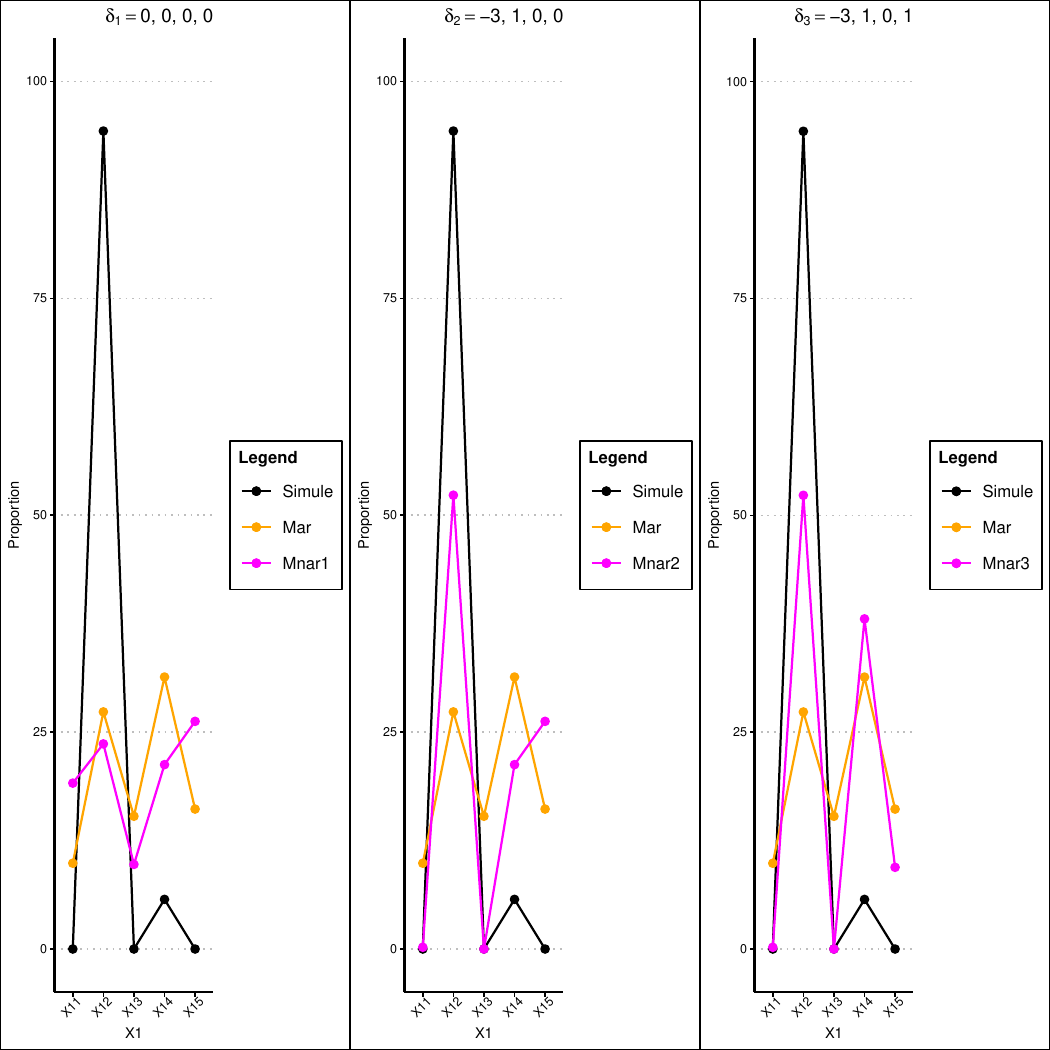}
   \caption{Non-hierarchical context (missing intermediate categories). Proportions of ordinal variable $X_1$ for $R = 0$ ($X_1$ missing) for 1000 simulations and 10 imputations with \texttt{R} package \texttt{mice}: simulated (Simule), MAR (Mar), MNAR (Mnar1 for $\delta_1 = (0, 0, 0, 0)$; Mnar2 for $\delta_2 = (-3, 1, 0, 0)$; Mnar3 for $\delta_3 = (-3, 1, 0, 1)$)  }
   \label{delta_no_hiercachic_int}
  \end{figure}

As $X_1$ has $(K=5)$ categories, $\delta$ is of length $(K-1 = 4)$. In Figure \ref{delta_no_hiercachic_int}, for the missing $X_1$ data, Mnar1, Mnar2, and Mnar3 represent varying degrees of deviation from MAR, derived respectively from $\delta_1 = (0, 0, 0, 0)$, $\delta_2 = (-3, 1, 0, 0)$, and $\delta_3 = (-3, 1, 0, 1)$. The proportions under MAR and those under Mnar1  $ (\delta_1 = (0, 0, 0, 0))$ appear to be almost similar. For Mnar2, unfortunately the category ($X_{14} = 4$) is not plausible given the missing data process. Finally for Mnar3 $(\delta_3 = (-3, 1, 0, 1))$, the average proportions are potentially consistent with the structure of $X_1$ without missing values. Indeed, the vector $\delta_3 = (-3, 1, 0, 1)$:

(i)  On average, minimize the proportions of the categories ($X_{11} = 1$, $X_{13} = 3$ and $X_{15} = 5$) and maximize those of intermediate categories ($X_{12} = 2$ and $X_{14} = 4$).

(ii) The proportion of $(X_{12} = 2)$ carries the highest weight compared to that of $(X_{14} = 4)$, consistent with the structure of the simulate ordinal variable $X_1$. 
From the above, it is possible to construct:

\begin{align*}
    \text{MNAR}_1 &= \begin{cases} 
        \text{Mnar1} & \text{if $X_1$ is missing} \\
        X_1 \; \text{(Observed)} & \text{if $X_1$ is not missing}
    \end{cases} \\[10pt]
   \text{MNAR}_2 &= \begin{cases} 
        \text{Mnar2} & \text{if $X_1$ is missing} \\
        X_1 \; \text{(Observed)} & \text{if $X_1$ is not missing}
    \end{cases} \\[10pt]
    \text{MNAR}_3 &= \begin{cases} 
        \text{Mnar3} & \text{if $X_1$ is missing} \\
        X_1 \; \text{(Observed)} & \text{if $X_1$ is not missing}
    \end{cases}
\end{align*}

\subsection{Relative biases, empirical standard deviation and  coverage rate.}
With the exception of simulated data without missing values, estimates under $\text{MNAR}_3$ offer an advantage. They are generally less biased than the MAR estimates, more precise than the MCAR estimates, and they also provide relatively better coverage compared to both situations (Table \ref{tab_no_hiercachic_int}).

  \begin{table}[H]
  \caption{Non-hierarchical context (missing intermediate categories). Relative bias, empirical standard deviation, and $95\%$ confidence interval coverage rate: SIMULATED  (Full data without missing values); CC (Completes cases analysis); MAR; MNAR ($\text{MNAR}_1$, $\text{MNAR}_2$ and $\text{MNAR}_3$)
}
  \centering
\begin{adjustbox}{width=\textwidth}
 \begin{tabular}{lccccccccccr}
 \rowcolor{gray!40}   &  \textbf{SIMULATED} & \textbf{CC} & \textbf{MAR} & $\textbf{MNAR}_1$ & $\textbf{MNAR}_2$ & $\textbf{MNAR}_3$   \\
 \textbf{Relative bias} ($\%$)     &&&&&&&\\
     intercept & -0.02  &3.21 & -3.34  &-4.64 & -2.61 &-0.86 \\
     $X_{12}$  &  0.92 &28.44 & 31.31 & 26.89 & 42.56 &27.71\\
      $X_{13}$ &  0.17 &-3.93  &-8.36 &-28.61 &-16.79 &-5.85\\
      $X_{14}$ & 0.72 &37.41  &31.24  &32.28  &17.89& 20.95\\
     $X_{15}$ & 0.88 &-2.78 &-11.64 &-10.82 &  1.28 & 2.00\\
   $X_{22}$ &  0.24&  0.35 &  7.49 &  8.40  & 4.83 & 2.88\\
   $X_{23}$ & -0.92 &-0.69 & 14.65 & 17.17 &  8.72 & 4.35 \\
   $X_{24}$ &0.18 & 0.29  & 7.47 &  8.36   &4.79 & 2.82\\

\rowcolor{gray!20}  \textbf{Empirical standard deviation}   &&&&&&\\
\rowcolor{gray!20} intercept &  0.12 &0.13 &0.12&  0.12 & 0.12  &0.12\\
 \rowcolor{gray!20}    $X_{12}$  &  0.15 &0.15 &0.14 & 0.13 & 0.14 & 0.15\\
 \rowcolor{gray!20}     $X_{13}$ & 0.15 &0.15 &0.13&  0.13  &0.14  &0.15 \\
 \rowcolor{gray!20}     $X_{14}$ & 0.14 &0.14& 0.14&  0.14&  0.14 & 0.14\\
 \rowcolor{gray!20}    $X_{15}$ &0.14 &0.14& 0.13 & 0.13 & 0.14 & 0.14 \\
\rowcolor{gray!20}   $X_{22}$ &  0.17 &0.19& 0.17  &0.17  &0.18 & 0.18\\
\rowcolor{gray!20}   $X_{23}$ & 0.16& 0.18 &0.17 & 0.16  &0.17 & 0.17 \\
\rowcolor{gray!20}   $X_{24}$ &0.17 &0.18 &0.18 & 0.17  &0.18  &0.18\\

\textbf{Coverage rate}    &&&&&&&\\
intercept &  0.95 & 0.95 & 0.94 &  0.93 &  0.94 &  0.95\\
     $X_{12}$  & 0.95 &0.53 &0.46  &0.58  &0.17 & 0.55\\
      $X_{13}$ & 0.94 &0.94 &0.93  &0.50 & 0.82  &0.93 \\
      $X_{14}$ & 0.95& 0.03& 0.11 & 0.09  &0.63 & 0.50\\
     $X_{15}$ & 0.95 &0.95 &0.68 & 0.73  &0.97 & 0.96\\
   $X_{22}$ &  0.94 &0.94 &0.86 & 0.84 & 0.92 & 0.94\\
   $X_{23}$ & 0.95 &0.94 &0.88 & 0.85 & 0.92  &0.95 \\
   $X_{24}$ & 0.95& 0.94 &0.86 & 0.83 & 0.92 & 0.94\\
 \hline
  \end{tabular}
 \end{adjustbox}
     \label{tab_no_hiercachic_int}
\end{table}%

\section{Non-hierarchical context results for continuous response (missing extreme categories)}\label{non:hierar:extrem:continuous}
We simulated a linear model, $\text{P}(Y = 1 \mid \mat{X}) = \mat{X}^\top \beta + \varepsilon$, where $Y$ is a continuous outcome variable, $\mat{X} = (X_1, X_2)$ is the set of independent variable is defined as in the previous section. The true parameter values are: \( \beta_0 = -1.5 \) for the intercept; \( \beta_{12} = 1 \), \( \beta_{13} = -2 \), \( \beta_{14} = 1.5 \), and \( \beta_{15} = 2 \) for \( X_1 \) (with category 1 as reference); and \( \beta_{22} = 2 \), \( \beta_{23} = 1 \), \( \beta_{24} = 2 \) for \( X_2 \) (with category 1 as reference). We generated 1,000 independent datasets of size \( n = 1000 \). The MNAR mechanism was as follows: if \( Y > 0 \), then 30\% of the values in category \( X_{11} \) were set to missing; if \( Y < 0 \), then 30\% of the values in category \( X_{15} \) were set to missing.

\subsection{Selection of sensitivity parameter vector.}
\begin{figure}[H]
    \centering
    \includegraphics[width=0.95\textwidth]{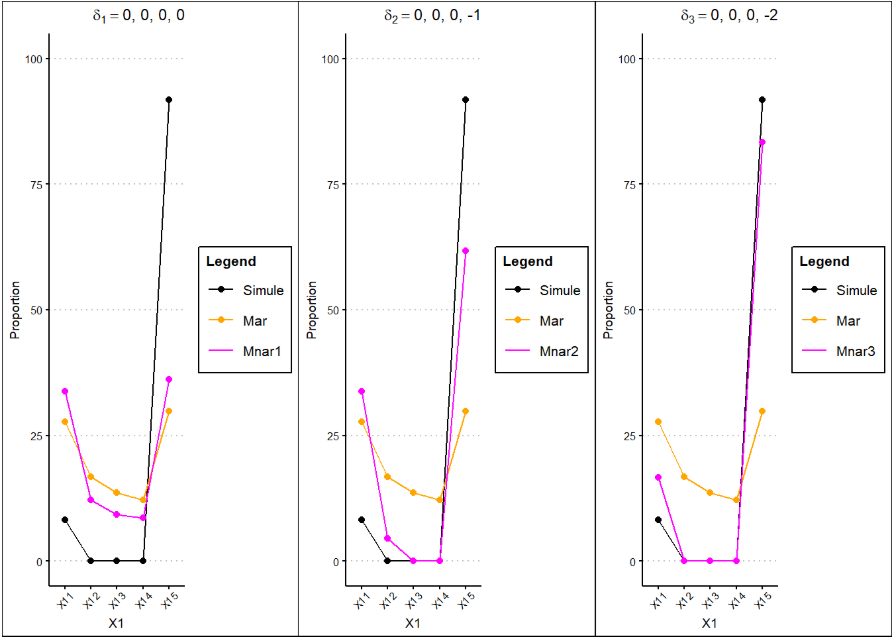}
   \caption{Non-hierarchical context results for continuous response (missing extreme categories). Proportions of ordinal variable $X_1$ for $R = 0$ ($X_1$ missing) for 1000 simulations and 20 imputations with \texttt{R} package \texttt{mice}: simulated (Simule), MAR (Mar), MNAR (Mnar1 for $\delta_1 = (0, 0, 0, 0)$; Mnar2 for $\delta_2 = (0, 0, 0, -1)$; Mnar3 for $\delta_3 = (0, 0, 0, -2)$)  }
   \label{non_hierachical_ext_cont}
  \end{figure}

As $X_1$ has $(K=5)$ categories, $\delta$ is of length $(K-1 = 4)$. In Figure \ref{non_hierachical_ext_cont}, for the missing $X_1$ data, Mnar1, Mnar2, and Mnar3 represent varying degrees of deviation from MAR, derived respectively from $\delta_1 = (0, 0, 0, 0)$, $\delta_2 = (0, 0, 0, -1)$, and $\delta_3 = (0, 0, 0, -2)$.  For the vector $(\delta_3 = (0, 0, 0, -2))$, the average proportions are potentially consistent with the structure of $X_1$ without missing values. We define the following MNAR-adjusted versions:

\begin{align*}
    \text{MNAR}_1 &= \begin{cases} 
        \text{Mnar1} & \text{if $X_1$ is missing} \\
        X_1 \; \text{(Observed)} & \text{if $X_1$ is not missing}
    \end{cases} \\[10pt]
    \text{MNAR}_2 &= \begin{cases} 
        \text{Mnar2} & \text{if $X_1$ is missing} \\
         X_1 \; \text{(Observed)} & \text{if $X_1$ is not missing}
    \end{cases} \\[10pt]
   \text{MNAR}_3 &= \begin{cases} 
        \text{Mnar3} & \text{if $X_1$ is missing} \\
         X_1 \; \text{(Observed)} & \text{if $X_1$ is not missing}
    \end{cases}
\end{align*}

\subsection{Relative biases, empirical standard deviation and  coverage rate.}
With the exception of simulated data without missing values, estimates under $\text{MNAR}_3$ offer an advantage. They are generally less biased than the MAR estimates, more precise than the MCAR estimates, and they also provide relatively better coverage compared to both situations (Table \ref{tab_no_hiercachic_cont}).

\vspace{0.5cm}

\begin{table}[H]
  \caption{Non-hierarchical context results for continuous response (missing extreme categories). Relative bias, empirical standard deviation, and $95\%$ confidence interval coverage rate: SIMULATED (Full data without missing values); CC (Completes cases analysis); MAR; MNAR ($\text{MNAR}_1$, $\text{MNAR}_2$, and $\text{MNAR}_3$)}
  \centering
\begin{adjustbox}{width=\textwidth}
 \begin{tabular}{lcccccc}
 \rowcolor{gray!40} & \textbf{SIMULATED} & \textbf{CC} & \textbf{MAR} & $\textbf{MNAR}_1$ & $\textbf{MNAR}_2$ & $\textbf{MNAR}_3$ \\
 \textbf{Relative bias} (\%) \\
Intercept & 0.23 & -0.94 & -2.11 & -2.28 & -2.43 & -1.07 \\
$X_{12}$  & 0.22 & 9.05  & 5.16  & 1.86  & -1.29 & 0.00  \\
$X_{13}$  & -0.11 & -2.12 & -8.38 & -7.27 & -3.42 & -1.19 \\
$X_{14}$  & -0.16 & 2.79  & -2.51 & -2.32 & -1.26 & -0.23 \\
$X_{15}$  & 0.09  & 0.54  & -4.28 & -2.64 & 1.07  & 0.33  \\
$X_{22}$  & 0.07  & -0.72 & -3.76 & -3.22 & -2.28 & -1.43 \\
$X_{23}$  & 0.55  & 2.28  & -3.45 & -1.46 & 0.53  & 0.71  \\
$X_{24}$  & 0.14  & -0.59 & -3.65 & -3.12 & -2.18 & -1.33 \\
\rowcolor{gray!20} \textbf{Empirical standard deviation} \\
\rowcolor{gray!20} Intercept & 0.07 & 0.07 & 0.07 & 0.07 & 0.07 & 0.07 \\
\rowcolor{gray!20} $X_{12}$  & 0.06 & 0.06 & 0.06 & 0.06 & 0.06 & 0.06 \\
\rowcolor{gray!20} $X_{13}$  & 0.08 & 0.08 & 0.08 & 0.08 & 0.08 & 0.08 \\
\rowcolor{gray!20} $X_{14}$  & 0.09 & 0.09 & 0.08 & 0.08 & 0.09 & 0.09 \\
\rowcolor{gray!20} $X_{15}$  & 0.09 & 0.09 & 0.09 & 0.09 & 0.09 & 0.09 \\
\rowcolor{gray!20} $X_{22}$  & 0.10 & 0.10 & 0.10 & 0.10 & 0.09 & 0.09 \\
\rowcolor{gray!20} $X_{23}$  & 0.09 & 0.09 & 0.09 & 0.09 & 0.09 & 0.09 \\
\rowcolor{gray!20} $X_{24}$  & 0.10 & 0.10 & 0.10 & 0.10 & 0.10 & 0.10 \\
\textbf{Coverage rate} \\
Intercept & 0.94 & 0.93 & 0.94 & 0.93 & 0.92 & 0.94 \\
$X_{12}$  & 0.96 & 0.72 & 0.90 & 0.96 & 0.97 & 0.97 \\
$X_{13}$  & 0.94 & 0.91 & 0.50 & 0.60 & 0.86 & 0.94 \\
$X_{14}$  & 0.95 & 0.90 & 0.94 & 0.94 & 0.94 & 0.95 \\
$X_{15}$  & 0.94 & 0.94 & 0.89 & 0.94 & 0.94 & 0.95 \\
$X_{22}$  & 0.95 & 0.95 & 0.90 & 0.91 & 0.93 & 0.95 \\
$X_{23}$  & 0.95 & 0.94 & 0.96 & 0.97 & 0.96 & 0.96 \\
$X_{24}$  & 0.94 & 0.93 & 0.89 & 0.91 & 0.93 & 0.93 \\
\hline
\end{tabular}
\end{adjustbox}
\label{tab_no_hiercachic_cont}
\end{table}

\end{document}